\documentstyle[amssymb,psfig]{mn2e}

\title[From young massive star cluster to old globular]{From young
massive star cluster to old globular: the $L_V-\sigma_0$ relationship
as a diagnostic tool}

\author[R. de Grijs, M.I. Wilkinson \& C.N. Tadhunter]{Richard de
Grijs,$^{1}$\thanks{E-mail: R.deGrijs@sheffield.ac.uk}\thanks{Guest
researcher at the Instituto Nacional de Astrof\'\i sica Optica y
Electr\'onica (INAOE), Luis Enrique Erro 1, Tonantzintla, Puebla
72840, Mexico} Mark I. Wilkinson,$^2$ and Clive N. Tadhunter$^1$ \\
$^1$ Department of Physics \& Astronomy, The University of Sheffield,
Hicks Building, Hounsfield Road, Sheffield S3 7RH \\ $^2$ Institute of
Astronomy, University of Cambridge, Madingley Road, Cambridge CB3 0HA
}

\date{Received date; accepted date}
\pubyear{2004}

\begin{document}
\maketitle

\begin{abstract}
We present a new analysis of the properties of the young massive star
clusters forming profusely in intense starburst environments, which
demonstrates that these objects are plausible progenitors of the old
globular clusters (GCs) seen abundantly in the Local Group. The method
is based on the tight relationship for old GCs between their $V$-band
luminosities, $L_V$, and (central) velocity dispersions, $\sigma_0$.
We improve the significance of the relationship by increasing the GC
sample size and find that its functional form, $L_V/L_\odot \propto
\sigma_0^{1.57 \pm 0.10}$ (km s$^{-1}$), is fully consistent with
previous determinations for smaller Galactic and M31 GC samples. The
tightness of the relationship for a GC sample drawn from environments
as diverse as those found in the Local Group implies that its origin
must be sought in intrinsic properties of the GC formation process
itself.\\ We evolve the luminosities of those young massive star
clusters (YMCs) in the local Universe which have velocity dispersion
measurements to an age of 12 Gyr, adopting a variety of IMF
descriptions, and find that most YMCs will evolve to loci close to, or
to slightly fainter luminosities than the improved GC relationship. In
the absence of significant external disturbances, this implies that
these objects may potentially survive to become old GC-type objects
over a Hubble time. The main advantage of our new method is its
simplicity. Where alternative methods, based on dynamical mass
estimates, require one to obtain accurate size estimates and to make
further assumptions, the only observables required here are the
system's velocity dispersion and luminosity. The most important factor
affecting the robustness of our conclusions is the adopted form of the
initial mass function. We use the results of $N$-body simulations to
confirm that dynamical evolution of the clusters does not
significantly alter our conclusions about the likelihood of individual
clusters surviving to late times. Finally, we find that our youngest
observed clusters are consistent with having evolved from a relation
of the form $L_V/L_\odot \propto \sigma_0^{2.1_{-0.4}^{+0.5}}$ (km
s$^{-1}$). This relation may actually correspond to the origin of the
GC fundamental plane.
\end{abstract}


\section{Introduction}
\label{intro.sec}

Young, massive star clusters (YMCs) are the most notable and
significant end products of violent star-forming episodes triggered by
galaxy collisions, mergers, and close encounters. Their contribution
to the total luminosity induced by such extreme conditions dominates,
by far, the overall energy output due to gravitationally-induced
star formation (e.g., Holtzman et al. 1992, Whitmore et al. 1993,
O'Connell et al. 1994, Conti et al. 1996, Watson et al. 1996, Carlson
et al. 1998, de Grijs et al.  2001, 2003a,b,c,d,e).

The question remains, however, whether or not at least a fraction of
the compact YMCs seen in abundance in extragalactic starbursts, are
potentially the progenitors of globular cluster (GC)-type objects. If
we could settle this issue convincingly, one way or the other, the
implications of such a result would have profound and far-reaching
implications for a wide range of astrophysical questions, including
(but not limited to) our understanding of the process of galaxy
formation and assembly, and the process and conditions required for
star (cluster) formation. Because of the lack of a statistically
significant sample of similar nearby objects, however, we need to
resort to either statistical arguments or to the painstaking approach
of case by case studies of individual objects in more distant galaxies.

The present state-of-the-art teaches us that the sizes, luminosities,
and -- in several cases -- spectroscopic mass estimates of most
(young, massive) extragalactic star cluster systems are fully
consistent with the expected properties of young Milky Way-type GC
progenitors (e.g., Meurer 1995, van den Bergh 1995, Ho \& Filippenko
1996a,b, Schweizer \& Seitzer 1998, de Grijs et al. 2001, 2003d).

However, the postulated evolutionary connection between the recently
formed YMCs in regions of violent star formation and starburst
galaxies, and old GCs similar to those in the Galaxy, M31, M87, and
other old elliptical galaxies is still a contentious issue. The
evolution and survivability of YMCs depend crucially on the stellar
initial mass function (IMF) of their constituent stars (cf. Smith \&
Gallagher 2001): if the IMF is too shallow, i.e., if the clusters are
significantly depleted in low-mass stars compared to (for instance)
the solar neighbourhood, they will disperse within a few orbital
periods around their host galaxy's centre, and most likely within
about a billion years of their formation (e.g., Gnedin \& Ostriker
1997, Goodwin 1997a, Smith \& Gallagher 2001, Mengel et al. 2002).

Ideally, one would need to obtain (i) high-resolution spectroscopy of
all clusters in a given cluster sample in order to obtain dynamical
mass estimates (we will assume, for the purpose of the present
discussion, that our YMCs are fully virialised based on their ages of
$\gtrsim 10^7$ yr, i.e., many crossing times old) and (ii)
high-resolution imaging (e.g., with the {\sl Hubble Space Telescope;
HST}) to measure their luminosities and sizes. 

In this paper, we explore the potential of a novel method to compare
the properties of YMCs in the context of those of old GC systems, and
predict their evolution over a Hubble time. In Section
\ref{diagnostic.sec} we outline the basic diagnostic tool we will use,
based on the distribution of old GCs in $L_V-\sigma_0$ space
(luminosity vs. central velocity dispersion). We extend this idea to
younger clusters in Section \ref{extension.sec}, and discuss the
uncertainties involved in our assumptions in Section
\ref{uncertainties.sec}. Section \ref{discussion.sec} provides a
detailed discussion of the implications of our results, and we
conclude the paper in Section \ref{summary.sec}.

\section{The $L_V-\sigma_0$ plane as a diagnostic tool for old globular 
clusters}
\label{diagnostic.sec}

\begin{figure*}
\psfig{figure=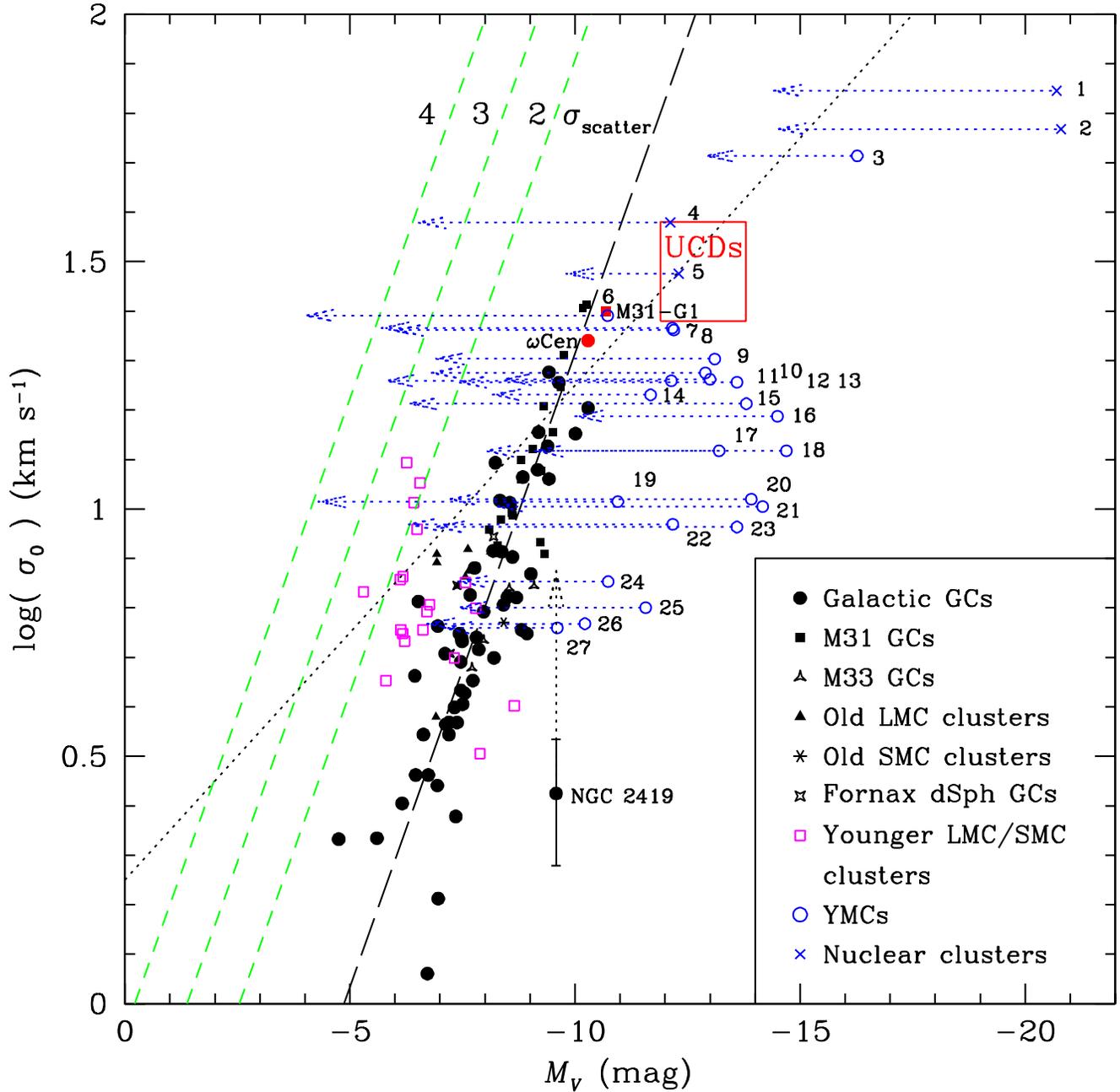,width=1.05\textwidth}
\caption[]{Diagnostic figure used to compare old GCs to YMCs with
(central) velocity dispersion measurements available in the
literature. The filled symbols correspond to the old GCs in the Local
Group, as indicated in the legend; the best-fitting relation for these
old clusters is shown by the long-dashed line. The short-dashed
(green) lines are displaced from this best-fitting relationship by,
respectively, 2, 3, and 4 times the scatter in the data points around
the best-fitting line, $\sigma_{\rm scatter}$, adopting a Gaussian
distribution of the scatter for simplicity. The dotted line
corresponds to the Faber-Jackson relationship for elliptical galaxies
(see text), which bisects the locus of the recently discovered
ultracompact dwarf galaxies (UCDs, in red; see Section
\ref{ymcdiagram.sec}). The numbered (blue) open circles are the
locations of the YMCs with measured velocity dispersions (see Table
\ref{clusids.tab} for the cluster IDs; the IDs are wherever possible
placed to the immediate right of the objects' locations in the
diagram, and in all other cases the ID labels follow the distribution
of the data points, e.g., as for clusters 7--8 and 10--13), which we
have evolved to a common age of 12 Gyr (represented by the blue dotted
arrows) using the GALEV SSP models for the appropriate metallicities
and ages of these objects (Table \ref{clusids.tab}). The (magenta)
open squares are the young compact clusters in the LMC and SMC (NGC
419). The most massive GCs in both the Galaxy and M31 ($\omega$ Cen
and G1, respectively) are also indicated (in red).}
\label{diagnostic1.fig}
\end{figure*}

It is well-known that the central velocity dispersion, $\sigma_0$, of
old GCs in the Galaxy and in M31 is tightly correlated with their
$V$-band luminosity, $M_V$ (e.g., Meylan \& Mayor 1986, Paturel \&
Garnier 1992, Djorgovski 1991, 1993, Djorgovski \& Meylan 1994,
Djorgovski et al. 1997). McLaughlin (2000a) suggests that this is a
consequence of the tighter relationship between a cluster's binding
energy, $E_{\rm b}$, and its luminosity, $E_{\rm b} \propto L^{2.05}$,
which is one of the defining relationships of the GC fundamental
plane. In Fig. \ref{diagnostic1.fig} we show this $L_V-\sigma_0$
relationship for old GCs, represented by the filled symbols. We not
only include the Galactic and M31 GCs (56 and 21 objects,
respectively; Pryor \& Meylan 1993, Djorgovski et al. 1997, Dubath \&
Grillmair 1997, Dubath, Meylan \& Mayor 1997; photometry from Crampton
et al. 1985, Bonoli et al. 1987, Reed, Harris \& Harris 1994), but
have also added -- for the first time -- the data points for the
($>10$ Gyr) old compact Magellanic Cloud clusters (9 clusters; Dubath
et al. 1993, 1997; photometry from Bica et al. 1996, de Freitas
Pacheco, Barbuy \& Idiart 1998), and the old GCs in M33 (Larsen et
al. 2002) and the Fornax dwarf spheroidal (dSph) galaxy (Dubath et
al. 1992, 1993) with available velocity dispersion measurements (4 and
3 GCs, respectively, for M33 and the Fornax dSph). Although
uncertainty estimates are available for both the photometry and the
central velocity dispersions, we decided not to include error bars for
the individual objects for reasons of clarity. As an example, slightly
larger than typical error bars are shown for NGC 2419; generally
speaking, the uncertainties in the central velocity dispersion are
$\lesssim 30$--40 per cent (or 0.10--0.15 dex), while the photometric
uncertainties are mostly smaller than the symbol sizes.

We find that the additional Local Group GCs follow, within the
measurement uncertainties, the $L_V-\sigma_0$ relationship for the
Galactic and M31 GCs. This is consistent with unpublished results for
the Fornax dSph and Small Magellanic Cloud (SMC) GCs referred to by
Djorgovski \& Meylan (1994).

The best-fitting relationship between the GC luminosities and their
central velocity dispersion is represented by the long-dashed line in
Fig. \ref{diagnostic1.fig}, which has the functional form
\begin{equation}
\sigma_0 (\mbox{km s}^{-1}) \propto \Bigl( \frac{L_V}{L_\odot}
\Bigr)^{0.64 \pm 0.04}
\end{equation}
or
\begin{equation}
\frac{L_V}{L_\odot} \propto \sigma_0^{1.57 \pm 0.10} (\mbox{km
s}^{-1}),
\end{equation}
with correlation coefficient $\Re = -0.817$, when expressed in the
logarithmic units used in Fig. \ref{diagnostic1.fig}. Based on his
identification of a GC fundamental plane, McLaughlin (2000a) predicted
a dependence of the form $\sigma_0 \propto (L/L_\odot)^{0.525}$ for
the pre-core collapse GCs in the Milky Way. He found that the form of
the correlations obtained by projecting the GC fundamental plane
depends only weakly on cluster properties such as Galactocentric
distance and concentration -- in fact, these affect the normalisations
of the relations rather than their slopes. Our larger data set
displays a relationship that is very similar to the predicted one.

The most discrepant data point among the old GCs is that of the
Galactic GC NGC 2419, as indicated in Fig. \ref{diagnostic1.fig}. It
is one of the most luminous Galactic GCs, and yet has one of the
lowest measured central velocity dispersions; both of these
observational parameters are well determined and the uncertainties are
too small to allow for the cluster to fall within the normal scatter
around the best-fitting relationship (cf. Olszewski, Pryor \& Schommer
1993). The arrow extending from the GC's location to higher velocity
dispersions indicates the expected value for its central velocity
dispersion based on its structural parameters and calculated using
single-mass isotropic King models with a constant mass-to-light (M/L)
ratio of $M/L_V = 3$ (Gnedin et al. 2002). NGC 2419 is a large
(half-light radius, $R_h \simeq 17.9$ pc), old ($\sim 12.3$ Gyr) outer
halo GC, located at a Galactocentric distance of $R_{\rm GC} \sim
91.5$ kpc (Harris 1996). It is possibly not a normal GC, but has been
speculated to be the stripped core of a former dSph galaxy (e.g., van
den Bergh \& Mackey 2004; but see Section \ref{ymcdiagram.sec}). Its
exclusion from the GC sample used to derive the best-fitting
relationship between $\sigma_0$ and $L_V$ does not alter this
relationship significantly.

The slope of the combined best-fitting relationship for all old Local
Group GCs with measured velocity dispersions is, within the
measurement uncertainties, consistent with the slopes most recently
determined by Djorgovski et al. (1997) for both the Galactic and M31
GCs individually ($1.7 \pm 0.3$ vs. $1.9 \pm 0.15$) and for the
combined Galactic/M31 GC sample ($1.7 \pm 0.15$). In
Fig. \ref{diagnostic1.fig} we have also indicated the 2, 3 and
$4\sigma$ envelopes toward fainter luminosities of the scatter of the
GC data points about the best-fitting relationship (short-dashed
lines; we have adopted a Gaussian distribution of the scatter, for
reasons of simplicity). We will return to these envelopes in Section
\ref{extension.sec}, where we will discuss the distribution and
evolution of the younger clusters also included in this figure, and
shown as the open circles and open squares.

\section{Extending the globular cluster idea}
\label{extension.sec}

\subsection{Understanding the input data set}
\label{inputdata.sec}

Encouraged by the tightness of the $L_V-\sigma_0$ relationship for old
Local Group GCs, we added the data points for the YMCs for which
velocity dispersion measurements are available in the literature.
These are indicated by the numbered open circles; Table
\ref{clusids.tab} provides an overview of the YMC identifications and
their age and metallicity measurements, and photometry. The YMCs are
ranked in order of decreasing (central) velocity dispersion. Since
most velocity dispersion measurements in the literature are given as
the ``observed'' velocity dispersion, corresponding to the
one-dimensional line-of-sight component, and denoted by $\sigma_{\rm
los}$ or $\sigma_x$, where relevant we corrected these measurements to
reflect the {\it central} value of the velocity dispersion profile. In
practice, this corresponds to applying an aperture correction to the
measurements from the effective size of the apertures used (typically
corresponding to $\sim 2-3 R_h$, for a given YMC). We adopted
Djorgovski et al.'s (1997) correction for M31 GCs of $\sigma_0 \simeq
1.14 \sigma_{\rm los}$ (see also McLaughlin 2000a). Although the exact
value of the clusters' concentration, $c$, is unknown in most cases,
this correction is applicable where $r_{\rm tidal} \gtrsim 3 r_{\rm
core}$ (so that $c \gtrsim 0.5$). This condition is met for all of the
YMCs in our sample.

Djorgovski et al. (1997) estimated the uncertainty of this correction
to be a few per cent, i.e. comparable to the measurement errors. We
note that while this procedure possibly introduces uncertainties that
are hard to quantify, our subsequent analysis is based on these values
in {\it logarithmic} parameter space, where the impact of these
uncertainties is minimised, $\lesssim \pm 0.05$ dex (see McLaughlin
2000a). Yet, since the central velocity dispersions defining the $L_V
- \sigma_0$ relationship span more than an order of magnitude, our
analysis of the relationship in logarithmic space does not penalise us
in terms of the resultant accuracy.

It is less straightforward to understand the effects of conversions of
the original photometric data to the $V$ band used to construct
Fig. \ref{diagnostic1.fig}. Yet, because of the relatively small
number of YMCs with measured (central) velocity dispersions, we
endeavoured to include as large a data set as possible in order to
increase the statistical relevance of the comparison done in this
paper. The penultimate column in Table \ref{clusids.tab} indicates
whether a given photometric entry was taken from the original
reference, or derived from the original data. In the following
sections, we will discuss our approach to these derivations on an
object-by-object basis. 

Our photometric conversion procedures are based on the following
general principles:
\begin{itemize}
\item Where we needed to adopt a distance modulus to a given YMC's
host galaxy, we used the most up-to-date values contained in the
HyperLeda database\footnote{http://leda.univ-lyon1.fr/}, except for
M82, where we adopted $m-M = 27.8$ based on Freedman et al.'s (1994)
Cepheid-based distance to the M81/M82/NGC 3077 group.

\item Conversions from a given passband to the $V$ band are age and
metallicity sensitive; we used the best available age and metallicity
estimates, together with the most up-to-date GALEV simple stellar
population (SSP) models (Schulz et al. 2002, Anders \&
Fritze--v. Alvensleben 2003), and assuming a Kroupa (2001; hereafter
Kroupa01) IMF, covering the mass range from 0.1 to 100 M$_\odot$ (see
Section \ref{mv.sec} for details). The Kroupa01 IMF is one of the
current best descriptions of the mass distribution of the stellar
populations in the solar neighbourhood. Below, we will also discuss
the impact of adopting this IMF on the uncertainties in our resulting,
converted $V$-band magnitudes.
\end{itemize}

\subsubsection{The NGC 1614 nuclear clusters}

Puxley \& Brand (1999) obtained high-resolution mid-infrared
spectroscopy of the two nuclear star clusters in NGC 1614, using the
Gemini 8m telescope. They calculated the objects' individual
bolometric luminosities to be $L_{\rm bol} = (1.5 \pm 0.3) \times
10^{11}$ and ($1.7 \pm 0.3) \times 10^{11}$ L$_{{\rm bol},\odot}$,
respectively. Using the appropriate bolometric correction for the Sun,
we derive $M_{V,{\rm NC1}} \simeq -20.7$ and $M_{V,{\rm NC2}} \simeq
-20.8$ mag, respectively. The uncertainties here are dominated by the
uncertainties in the original conversion from mid-infrared flux to
bolometric luminosity. The combination of using the bolometric
correction for the Sun and a metallicity of 2 Z$_\odot$ contributes an
uncertainty of up to $\sim 0.15$ mag. This is of a similar order as
the uncertainties in the original photometry, where given in Table
\ref{clusids.tab}.

\subsubsection{The nuclear cluster in NGC 1042}

Photometry of the nuclear cluster in NGC 1042 was published by B\"oker
et al. (2004) and Walcher et al. (2004) as $M_{I,{\rm NC}} = -13.14$
mag. For the best age estimate of $\sim 10^9$ yr, our GALEV models for
the appropriate metallicity indicate $(V-I) \simeq 0.83$ mag, thus
leading to $M_{V,{\rm NC}} \simeq -12.3$ mag. The uncertainties in
this conversion owing to the IMF parameterisation adopted are minimal;
comparing the $(V-I)$ values for all of the IMFs discussed in Section
\ref{mv.sec} below, and assuming solar metallicity (see Table
\ref{clusids.tab}), we find a maximum difference among the $(V-I)$
colours predicted of $\Delta(V-I)_{\rm IMF} \lesssim 0.1$ mag, ranging
from $(V-I) = 0.79$ mag for the Salpeter (1955) IMF truncated at 1
M$_\odot$, to $(V-I) = 0.88$ mag for a non-truncated Salpeter IMF. By
having adopted the Kroupa01 IMF, we have essentially halved this
uncertainty.

A more important contribution to the photometric uncertainty arises
from the fact that we have assumed the NGC 1042 NC to behave as a
clean SSP. However, we note that this is perhaps a questionable
assumption: nuclear clusters are more likely to be contaminated by
secondary and tertiary star-formation episodes than more isolated star
clusters in the outer regions of their host galaxies (e.g., Cid
Fernandes et al. 2004), so that in essence we are measuring the
properties of a luminosity-weighted mean stellar population in this
case. We will return to this discussion below.

\subsubsection{YMCs in the Antennae galaxies}
\label{antunc.sec}

Of the YMCs in the Antennae galaxies, only clusters [WS95]355 and
[M03] required photometric conversions to the $V$ band; for the other
YMCs we adopted the original photometry. Because of their young ages,
of $8.5 \pm 0.3$ and $8.0 \pm 0.3$ Myr, the photometric uncertainties
in the conversions to the $V$ band are more significant for these
clusters than for the older nuclear cluster in NGC 1042.

For [WS95]355, Mengel et al. (2002) reported only an upper limit in
the $V$ band, but a well-determined flux in $I$. It is in this age
range where uncertainties in the treatment of the more massive
component of any SSP, and in particular that of the red supergiants,
render colour transformations significantly uncertain. Adopting the
same set of IMFs as above, we find that $\Delta (V-I)_{\rm max,IMF}
\simeq 1.10$ mag, ranging from $(V-I) = 0.35$ mag for the Kroupa01 and
Kroupa, Tout \& Gilmore (1993, KTG93) IMFs to $(V-I) = 1.35$ mag for
the truncated Salpeter IMF. As we will see in Section \ref{mv.sec}
below, when evolved to an age of 12 Gyr, this YMC does stand out, by
$\Delta M_V \gg 1$ mag, from the majority of the other YMCs in our
sample. Therefore, we believe that we can confidently include this
object in our statistical analysis of the $L_V - \sigma_0$ diagnostic
diagram, despite this large photometric uncertainty, and despite the
considerable uncertainty introduced by the poorly bracketed effects of
internal extinction in the Antennae system (see Sect. \ref{mv.sec}
below).

Unfortunately, we cannot be as confident for cluster [M03]. For this
object, our only photometric data consists of the combination of a
dynamical mass estimate ($M_{\rm dyn} = (0.85 \pm 0.2) \times 10^6$
M$_\odot$) and a $K$-band M/L ratio of $\log(L_K/M) = 1.49$ (Mengel
2003). Using $M_{K,\odot} = 3.33$, we then obtain $M_{K,{\rm YMC}} =
-15.22$. Similar analysis as presented in the previous paragraph shows
that the inherent photometric uncertainties at its young age caused by
IMF variations amount to $\Delta (V-K) \simeq 1.15$ mag, ranging from
$(V-K) = 1.15$ mag for the KTG93 IMF to $(V-K) = 2.30$ mag for the
truncated Salpeter IMF. By adopting the Kroupa01 IMF as our IMF
parameterisation, we reduce this uncertainty to $\Delta (V-K) \simeq
0.95$ mag. Contrary to [WS95]355, [M03] does not stand out from the
sample objects in any specific way, and in view of the large
photometric uncertainty, we can only conclude that this cluster
appears to follow the trend set by the bulk of the sample (see Section
\ref{mv.sec}).

\subsubsection{The NGC 1487 YMCs}

Our $V$-band magnitudes for the three YMCs in NGC 1487, also observed
by Mengel (2003), were obtained using exactly the same procedure as
used for Antennae YMC [M03]. Once again, because of the YMCs' ages
clustering around 8 Myr, the photometric uncertainty owing to the
$K$-to-$V$ conversion is significant and highly IMF dependent, with
the most likely uncertainty on the order of $\Delta (V-K) \simeq 0.9$
mag, as discussed above. As we will see in Section \ref{mv.sec},
although these three objects show tentative differences with respect
to the majority of our cluster sample, when evolved to a common age of
12 Gyr, the large photometric uncertainty does not allow us to draw
firm conclusions on these perceived differences.

\subsubsection{YMCs in M82}

Of the three sample YMCs drawn from the large cluster sample in M82,
we used the original photometry of Smith \& Gallagher (2001) for
M82-F, which the authors attempted to correct for the effects of a few
saturated pixels. Nevertheless, we are more confident using the
corrected $V$ magnitude (the quoted uncertainty in which already
includes the effects caused by the saturated pixels) than McCrady et
al.'s (2003) near-infrared {\sl HST} photometry, in view of the much
larger uncertainties introduced by filter conversions using a given
IMF (see above). McCrady et al. (2005) report new ACS observations of
M82-F in the {\sl HST} F555W band, but do not give the cluster's
integrated magnitude in this filter. In view of the uncertainties
involved in converting their F814W luminosity to a $V$-band flux, we
are hesitant to take this approach.

For objects MGG-9 and 11, we have to resort to a similar technique as
applied to the NGC 1487 clusters and to YMC [M03] in the Antennae
galaxies. McCrady et al. (2003) provide {\sl HST}-equivalent $H$
(F160W) and $K'$-band (F222M) photometry for these two objects. Given
their age of $\sim 7-12$ Myr, the uncertainty due to the passband
conversion amounts to $\Delta (V-m_{\rm F160W}) \simeq 0.7$ mag for
the same range of IMF parameterisations as used above. In addition, as
we will show below (Section \ref{mv.sec}), the additional photometric
uncertainties owing to the intrinsic uncertainties in the F160W-band
extinction estimates of McCrady et al. (2003) are considerable.

\subsubsection{Concluding remarks}

Based on the analysis of the effects of passband conversions on the
quality of the input photometry for the diagnostic $L_V-\sigma_0$
diagram, we conclude that the resulting uncertainties are most
significant for the youngest objects. These converted $V$-band
magnitudes should therefore be treated with caution. In our sample of
27 YMCs, this affects six objects, for which $\Delta M_V \lesssim 1$
mag. For the remainder of the sample, the photometric uncertainties in
the input data are significantly smaller, and mostly on the order of
up to several tenths of a magnitude.

\begin{table*}
\caption[ ]{\label{clusids.tab}Cluster IDs, age and metallicity estimates.}
{\scriptsize
\center{
\begin{tabular}{rllclcccc}
\hline
\hline
\multicolumn{1}{c}{ID} & \multicolumn{1}{c}{Cluster$^a$} &
\multicolumn{1}{c}{Age (yr)} & \multicolumn{1}{c}{Ref.} &
\multicolumn{1}{c}{Metallicity} & \multicolumn{1}{c}{Ref.} &
\multicolumn{1}{c}{$M_V$} & \multicolumn{1}{c}{Original /} &
\multicolumn{1}{c}{Original} \\
& & & & & & \multicolumn{1}{c}{adopted} & \multicolumn{1}{c}{Derived}
& \multicolumn{1}{c}{Ref.}\\
\hline
 1 & NGC 1614-NC1                & $(6-8)       \hfill \times 10^6$ & 25 & $2 Z_\odot$   &  2         & $-20.7$  & D & 25 \\
 2 & NGC 1614-NC2                & $(6-8)       \hfill \times 10^6$ & 25 & $2 Z_\odot$   &  2         & $-20.8$  & D & 25 \\
 3 & NGC 7252-W3                 & $3.0         \hfill \times 10^8$ & 18 & $0.5 Z_\odot$ & 18         & $-16.27 \pm 0.02$ & O & 17,18, \\
   &                             & $(5.4\pm0.2) \hfill \times 10^8$ & 26 &               &            &          &   & 26 \\
 4 & IC 342-NC                   & $10^{6.8-7.8}$                   &  5 & $\gtrsim 2 Z_\odot$ & 30   & $-12.12$ & O &  5 \\
   &                             &                                  &    & $^d$          & 28         \\
 5 & NGC 1042-NC                 & $\hfill 10^9$                    & 31 & $Z_\odot$     & $^e$       & $-12.3$  & D &  6,31 \\
 6 & Antennae-$\mbox{[WS95]355}$ & $(8.5\pm0.3) \hfill \times 10^6$ & 21 & $Z_\odot$     & 21         & $-10.72$ & D & 21 \\
 7 & Antennae-$\mbox{[W99]15}$   & $(8.7\pm0.3) \hfill \times 10^6$ & 21 & $Z_\odot$     & 21         & $-12.16$ & O & 21 \\
 8 & NGC 1487-3                  & $(7.9\pm0.5) \hfill \times 10^6$ & 22 & $0.15-0.4 Z_\odot$ & 1$^f$ & $-12.2$  & D & 22 \\
 9 & NGC 1487-1                  & $(8.1\pm0.5) \hfill \times 10^6$ & 22 & $0.15-0.4 Z_\odot$ & 1$^f$ & $-13.1$  & D & 22 \\
10 & NGC 1487-2                  & $(8.5\pm0.5) \hfill \times 10^6$ & 22 & $0.15-0.4 Z_\odot$ & 1$^f$ & $-12.9$  & D & 22 \\
11 & Antennae-$\mbox{[W99]16}$   & $(10\pm2)    \hfill \times 10^6$ & 21 & $Z_\odot$     & 21         & $-12.14$ & O & 21 \\
12 & M82 MGG-9                   & $10^{+2}_{-3}\hfill \times 10^6$ & 19 & $Z_\odot$  & 19         & $-15.1$  & D & 19$^i$ \\
13 & NGC 1569-A1$^b$             & $(4-5)       \hfill \times 10^6$ & 12 & [Fe/H] $=-0.7$& 3,8,       & $-13.6$  & O &  7$^j$ \\
   &                             & $(7-10)      \hfill \times 10^6$ & 12,24 &            & 10,13      \\
   &                             & $(12\pm4)    \hfill \times 10^6$ & 4$^c$ & [Fe/H] $=-1.7$ & 4$^c$  \\
14 & NGC 4214-13                 & $(2.0\pm0.4) \hfill \times 10^8$ & 15 & $0.4 Z_\odot$ & 15         & $-11.68$ & O & 15 \\
15 & Antennae-$\mbox{[W99]2}$    & $(6.6\pm0.3) \hfill \times 10^6$ & 21 & $2 Z_\odot$   & 21         & $-13.81$ & O & 21 \\
16 & M82-F                       & $(60\pm20)   \hfill \times 10^6$ & 27 & $Z_\odot$     & 27         & $-14.5 \pm 0.3$ & O & 27$^i$ \\
   &                             & $(40-60)     \hfill \times 10^6$ & 19,20 \\
17 & M82 MGG-11                  & $9^{+3}_{-2} \hfill \times 10^6$ & 19 & $Z_\odot$     & 19         & $-14.5$  & D & 19$^i$ \\
18 & NGC 1705-I                  & $(10-20)     \hfill \times 10^6$ & 11 & $0.5 Z_\odot$ & 29         & $-14.7$  & O & 23 \\
   &                             & $12^{+3}_{-1}\hfill \times 10^6$ & 29 \\
19 & Antennae-$\mbox{[WS95]331}$ & $(8.1\pm0.3) \hfill \times 10^6$ & 21 & $Z_\odot$     & 21,22      & $-10.95 \pm 0.08$ & O & 22 \\
20 & Antennae-$\mbox{[W99]1}$    & $(8.1\pm0.5) \hfill \times 10^6$ & 21 & $Z_\odot$     & 21         & $-13.92$ & O & 21 \\
21 & NGC 6946-1447               & $(15\pm5)    \hfill \times 10^6$ & 14 & $Z_\odot$     & 9,14       & $-14.17$ & O & 15 \\
   &                             & $(12-13)     \hfill \times 10^6$ &  9 & $^g$          & 28         \\
   &                             & $11^{+2}_{-3}\hfill \times 10^6$ & 15 \\
22 & NGC 5236-805                & $13^{+7}_{-5}\hfill \times 10^6$ & 16 & $^h$          & 16         & $-12.17 \pm 0.37$ & O & 16 \\
23 & Antennae-$\mbox{[M03]}$     & $(8.0\pm0.3) \hfill \times 10^6$ & 22 & $Z_\odot$     & 21,22      & $-13.6$  & D & 22 \\
24 & NGC 4449-47                 & $2.8^{+0.7}_{-0.6}\hfill\times 10^8$ & 15 & $0.4 Z_\odot$ & 15     & $-10.74$ & O & 15 \\
25 & NGC 5236-502                & $(1.0\pm0.2) \hfill \times 10^8$ & 16 & $^h$          & 16         & $-11.57 \pm 0.15$ & O & 16 \\
26 & NGC 4214-10                 & $(2.0\pm0.4) \hfill \times 10^8$ & 15 & $0.4 Z_\odot$ & 15         & $-10.22$ & O & 15 \\
27 & NGC 4449-27                 & $7.9^{+6.2}_{-3.5}\hfill\times 10^8$ & 15 & $0.4 Z_\odot$ & 15     & $-9.61$  & O & 15 \\
\hline
\end{tabular}
}
\flushleft
{\sc Notes}: $^a$ ``NC'' refers to nuclear clusters; The original
Antennae cluster data is from Whitmore \& Schweizer (1995; [WS95]),
Whitmore et al. (1999; [W99]) and Mengel (2003; [M03]); $^b$ We
adopted an age of 8 Myr for this cluster; $^c$ Based on broad-band
photometry; $^d$ 12 + log(O/H) $\sim 9.3$ at a radius of 4 kpc and
rising inward; $^e$ Although no metallicity estimates are available,
we adopted solar metallicity on the basis that the cluster was likely
formed from pre-enriched material; $^f$ They adopted $0.25 Z_\odot$;
$^g$ 12 + log(O/H) $\sim 9.15$ in the galactic centre; $^h$ $Z = 0.4
Z_\odot, Z_\odot$ and $2.5 Z_\odot$ all give similar results; we
adopted solar metallicity; $^i$ These absolute magnitudes were
corrected for extinction by the original authors, so that they
represent $M_V^0$; $^j$ Based on the absolute magnitude in the {\sl
HST} F555W filter.
{\bf References}: 1, Ag\"uero \& Paolantonio (1997); 2, Aitken et
al. (1981); 3, Aloisi et al. (2001); 4, Anders et al. (2004); 5,
B\"oker et al. (1999); 6, B\"oker et al. (2005); 7, De Marchi et
al. (1997); 8, Devost et al. (1997); 9, Efremov et al. (2002); 10,
Greggio et al. (1998); 11, Ho \& Filippenko (1996b); 12, Hunter et
al. (2000); 13, Kobulnicky \& Skillman (1997); 14, Larsen et
al. (2001); 15, Larsen et al. (2004); 16, Larsen \& Richtler (2004);
17, Maraston et al. (2001); 18, Maraston et al. (2004); 19, McCrady et
al. (2003); 20, McCrady et al. (2005); 21, Mengel et al. (2002); 22,
Mengel (2003); 23, O'Connell et al. (1994); 24, Origlia et al. (2001);
25, Puxley \& Brand (1999); 26, Schweizer \& Seitzer (1998); 27, Smith
\& Gallagher (2001); 28, Tosi \& D\'\i az (1985); 29, V\'azquez et
al. (2004); 30, Verma et al. (2003); 31, Walcher et al. (2004).\hfill}
\end{table*}

\subsection{A diagnostic diagram for testing the universality of the YMC 
formation process?}
\label{ymcdiagram.sec}

In order to compare the YMC loci with those of the GCs, we evolved the
YMC luminosities to a common age of 12 Gyr (see the dotted arrows
toward fainter luminosities in Fig. \ref{diagnostic1.fig}), using the
most recent GALEV SSP models, and assuming a ``standard'' Salpeter
IMF, covering the mass range from 0.1 to 100 M$_\odot$. We took
special care to adopt the most appropriate SSP models, based on their
current age and metallicity (see Table \ref{clusids.tab}). In the
remainder of this paper, wherever we refer to the evolution of our YMC
sample to an age of 12 Gyr, we implicitly assume this standard
Salpeter IMF, and stellar evolution following the GALEV SSPs, unless
indicated otherwise.

At first sight, we identify three main results based on this
photometric evolution:
\begin{enumerate}
\item Almost all YMCs appear to evolve to loci on the fainter side of
the old GC relationship. This may give us a handle on the functional
form of the realistic IMF, if we assume that these YMCs will evolve to
obey the GC $L_V-\sigma_0$ relationship at old age. In addition, it
may help us to determine whether the YMC formation process itself is
(close to) universal;
\item For most YMCs, luminosity evolution governed by a Salpeter-type
IMF results in these objects ending up very close to the best-fitting
GC relationship by the time they reach an age of 12 Gyr;
\item A small fraction ($\lesssim 30$ per cent) of the YMCs appear to
form a distinct group at significantly fainter luminosities than
expected for old GC-type objects, if we evolve their luminosities
assuming a Salpeter-type IMF. This implies that if their {\it initial}
mass function (MF) was similar to the Salpeter law, their {\it
present-day} MF must be significantly depleted in low-mass stars if
they are assumed to evolve to the GC relationship, as we will see
below. Alternatively, if the IMF was unlike a Salpeter-type IMF, then
comparison with the clusters discussed in point (ii) would suggest
that IMF variations exist in the highest-density regions in active
starbursts, the birth places of these YMCs. In this context, it is
worth noting that the tightness of the $L_V-\sigma_0$ relationship for
the Local Group GCs, and the lack of any significant dependence of GC
properties on metallicity (see also Sect. \ref{uncertainties.sec} and
McLaughlin 2000b) points to a universal IMF in -- at least -- the
Local Group.
\end{enumerate}

Of the 20 YMCs with projected central velocity dispersions smaller
than those of the most massive GC candidates in the Local Group
($\omega$Cen in the Galaxy, and G1--Mayall II in M31) 13 objects have
the potential to evolve to a position in the $L_V-\sigma_0$ diagnostic
diagram within $2\sigma_{\rm scatter}$ of the best-fitting GC
relationship. Since {\it all} of the GCs in our Local Group GC sample
fall well within this $2\sigma_{\rm scatter}$ envelope, we adopt this
envelope as the stability boundary for a cluster to survive for a
Hubble time (we realise that this is, of course, a relatively
arbitrary assumption, but we will use it simply to guide the
discussion). Of the remaining 7 YMCs with projected central velocity
dispersions smaller than those of $\omega$Cen and G1, 5 objects
overshoot even the $3\sigma_{\rm scatter}$ envelope if we adopt a
standard Salpeter IMF for their stellar content. If this IMF
assumption is valid, then these objects would appear to be too
dynamically hot, given their luminosities, to become old GC
counterparts. If they are to evolve to loci close to the
well-established GC relationship, their IMF (or their present-day MF)
must be significantly different from Salpeter; we will return to this
issue in Section \ref{mv.sec}.

The five objects with the largest projected central velocity
dispersions are suspected to be either nuclear star clusters, or
perhaps stripped dSph or dwarf elliptical (dE) nuclei (cf. NGC 7252-W3
= object 3; Maraston et al. 2004). Their range of central velocity
dispersions overlaps that of the recently discovered ``ultracompact
dwarf galaxies'' (UCDs) in the Fornax cluster (e.g., Hilker et
al. 1999, Drinkwater et al. 2000, 2003). The nature of these latter
objects is as yet unclear: they may be very large star clusters
(perhaps stripped nuclear clusters), or instead extremely compact dE
galaxies, such as M32. On the assumption that these objects constitute
a new class of galaxies, Drinkwater et al. (2003) argued that they
follow the Faber-Jackson (FJ) relation for elliptical galaxies, which
has a slope that is markedly different from that of the GC
relationship. The FJ relation for elliptical galaxies, and the loci of
the Fornax UCDs are also indicated in
Fig. \ref{diagnostic1.fig}. Intriguingly, the crossing point between
the FJ and GC relationships is very close to the locations of
$\omega$Cen and M31-G1 in the diagnostic diagram of
Fig. \ref{diagnostic1.fig}; both objects have been suggested to be the
stripped nuclei of dwarf galaxies captured by their host galaxies.

Unfortunately, however, neither the location by itself of the Fornax
UCDs on the FJ relationship, nor of any of the other (nuclear) star
clusters, provides conclusive evidence as to the nature of these
extremely massive objects, unless their dominant stellar populations
are older than $\sim 10-12$ Gyr. For the Fornax UCDs to evolve to the
GC relationship, their dominant stellar populations need only be as
young as (or younger than) $\sim 1.3 - 1.5$ Gyr, somewhat depending on
metallicity, again assuming that they are governed by a standard
Salpeter-type IMF and stellar evolution as described by the GALEV SSP
models.

Hilker et al. (1999) analysed two of the five Fornax UCDs in more
detail spectroscopically, and concluded that while object CGF 5-4 is
most likely older than $\sim 12$ Gyr (ages as young as 3 Gyr can be
excluded with confidence), the location of object CGF 1-4 in the
Mg$_2$ vs. $\langle$Fe$\rangle$ diagram suggests an age as young as
$3.0 \pm 1.5$ Gyr ($1\sigma$ uncertainty), based on its H$\beta$ line
strength. In addition, Drinkwater et al. (2000) point out that the
spectra of these objects are best fit by K-type stellar templates,
consistent with an old (metal-rich) stellar population.  This suggests
that they might be related to GCs, since dE galaxies observed with the
same set up are best fit by younger F and early G-type
templates. Thus, the nature of these intriguing objects is still an
open issue.

If we now consider our sample objects with the largest central
velocity dispersions in this context, and evolve their dominant
stellar populations to a common age of 12 Gyr, we find that they tend
toward the best-fitting GC line, although within the uncertainties
(see Section \ref{uncertainties.sec}) they are also consistent with
objects following the FJ relationship. We also note that while we have
used SSP models to evolve the luminosities of these nuclear clusters
to old age, this is strictly speaking not correct. Nuclear clusters
are not well described by ``simple'' stellar populations, but exhibit
(sometimes significant) age ranges (e.g., Cid Fernandes et
al. 2004). The implication of this is that, in fact, we may have {\it
overestimated} the lengths of the luminosity evolution arrows in
Fig. \ref{diagnostic1.fig} for these objects, depending on how much
their stellar contents deviate from the SSP approximation, and from a
Salpeter-type IMF (see Section \ref{mv.sec}). The main consequence of
this is that these nuclear clusters may indeed follow the FJ
relationship if they are able to survive to old age.

Thus, by placing the recently discovered UCDs in this context, we
believe that they may be closely related to nuclear star clusters, and
perhaps are the stripped nuclei of dE galaxies, akin to $\omega$Cen,
M31-G1, and NGC 7252-W3 (Maraston et al. 2004; see also Drinkwater et
al. 2003).

Let us now briefly return to the suggestion by van den Bergh \& Mackey
(2004) that the unusual GC NGC 2419 may also be a similar type of
object. If this were the case, we would expect the cluster to be
located close to either the FJ relation in Fig. \ref{diagnostic1.fig}
or -- if it were a genuine GC -- to the fundamental plane correlation
for Galactic GCs (e.g., Dubath et al. 1997, their Fig. 16, McLaughlin
2000a). In either case, the location of NGC 2419 is, respectively,
$\gtrsim 6\sigma$ and $\gtrsim 3\sigma$ (where $\sigma$ represents the
measurement uncertainty) removed from the fiducial
relationship. Therefore, we conclude that it is unlikely that NGC 2419
is the stripped core of a dSph galaxy.

We note that, thus far, we have only considered the evolution of the
YMCs in terms of their luminosity, and have ignored the possibility of
significant evolution of the central velocity dispersion over a Hubble
time. Following an initial phase of mass loss caused by stellar
evolution, the long-term dynamical evolution of star clusters is
dominated by evaporation due to internal relaxation and stripping due
to external, tidal shocks. The latter process removes mass (and
luminosity), but should not significantly affect the central velocity
dispersion (e.g., Djorgovski 1991, 1993, Djorgovski \& Meylan
1994). It is unclear, however, how the central velocity dispersion
evolves over time as a result of internal evolution in the presence of
external tidal fields, significant binary fractions, the effects of
mass segregation and core collapse. $N$-body simulations present an
ideal way to investigate this problem. However, despite the vast
literature on $N$-body simulations of star clusters, we are not aware
of any paper which presents the evolution of the central, projected
velocity dispersion of the simulated clusters. Therefore, in Section
\ref{sigma.sec} we investigate the evolution of the observable
properties of a set of simulated $N$-body clusters in order to
constrain the expected evolution of the observed $\sigma_0$.

\subsection{Comparison with previous predictions}
\label{prevpred.sec}

In the previous sections we have constructed a diagnostic tool that
could potentially tell us whether a given YMC might evolve into a
GC-type object over a Hubble time, based on only two observables: the
cluster's (central) velocity dispersion and its $V$-band luminosity
(or absolute magnitude). This provides a simpler and potentially more
reliable method to predict, to first order, the evolutionary fate of
YMCs than existing methods. In particular, the most common method to
assess this issue is based on the comparison of dynamical cluster mass
estimates with a variety of IMF descriptions in the (Age vs. M/L
ratio) plane. This method introduces two complications that we can in
principle avoid using the $L_V-\sigma_0$ approach: in order to
estimate an object's dynamical mass, one needs to (i) assume that the
virial theorem applies (which is generally assumed to hold for
clusters older than $\sim 10$ Myr), and (ii) obtain a reliable
measurement of the cluster radius. While the complication introduced
by the assumption of virialisation is minimal (although it may play a
significant role for the youngest objects in our sample!), measuring
reliable cluster radii is problematic for all but the nearest
objects. In addition, using the half-light radius as an estimate of
the volume occupied by the cluster implicitly assumes that the M/L
ratio is constant across the cluster -- an assumption that may be
unjustified in the presence of significant mass segregation, as shown
observationally (see, e.g., de Grijs et al. 2002b, and references
therein; see also Section \ref{m82clus.sec} below and the discussion
in McCrady et al. 2005). Thus, here we have presented a simpler and
potentially more reliable method to predict the approximate evolution
for a given YMC than currently available.

We will now compare the predictions from this new method to those
obtained from the dynamical mass estimates, in order to assess the
robustness of the $L_V-\sigma_0$ approach, on a case by case basis,
for those of our sample clusters for which this information is
available. Where appropriate, we will also point out those cases where
discrepancies between our new results and previous predictions occur;
these provide a useful insight into the uncertainties inherent to the
use of any of the methods currently employed in this field. For the
purposes of this discussion, we will consider whether the
observational data are consistent with the assumption that all
surviving old star clusters will obey the Local Group GC correlation
between $L_V$ and $\sigma_0$, within the uncertainties.

\subsubsection{Antennae clusters}
\label{antennae.sec}

Mengel et al. (2002) concluded, aided by ground-based $K$-band
luminosities, that clusters [W99]1 and [W99]2 appeared to have a
deficit of low-mass stars (see their Fig. 7), either because of a
shallower-than-Salpeter IMF slope down to stellar masses of $\sim 0.1$
M$_\odot$, or because of a low-mass IMF cut-off. Their results for
YMCs [W99]15, 16 and [WS95]355 are more consistent with a steeper IMF
slope, similar to or steeper than the standard Salpeter slope (or,
alternatively, an overabundance of low-mass stars compared to the
standard Salpeter IMF), down to low masses. These results are
supported by their {\sl HST}-based $V$-band observations for [W99]1,
15 and 16 (although the uncertainties for cluster [W99]1 make it a
potential object with a Salpeter-type slope; see their Fig. 6),
although the opposite trend is found for object [W99]2, at a level of
2--3 times the uncertainty in the measurements. This object appears to
be characterised by a decidedly larger proportion of low-mass stars
based on its $V$-band photometry than seemed to be the case based on
the $K$-band data (see below for a discussion). It is striking that
they seem to find systematically steeper IMF slopes (or, equivalently,
IMFs richer in low-mass stars) in the higher-density overlap region
between the two merging galaxies (containing clusters [W99]15, 16 and
[WS95]355; although [W99]16 may not be located in the densest region,
we believe its ambient density to be much higher than that in the
outer regions of the system; see also Mengel et al. [2002]), while the
low-mass deficient IMFs are found in the outer spiral arms (containing
objects [W99]1 and 2). Mengel (2003) obtained similar quality
measurements for the additional YMCs [M03] and [W99]331, both of which
appear to be characterised by a ``normal'' IMF with a Salpeter-type
slope down to 0.1 M$_\odot$ in their diagnostic (Age vs. $M/L_K$)
diagram.

If we adopt the assumption that these YMCs will eventually evolve to
loci close to the $L_V-\sigma_0$ relation for old GCs -- at least, if
they survive sufficiently long -- then our diagnostic $L_V-\sigma_0$
diagram suggests that clusters [WS95]331, [WS95]355, [W99]15, and
[W99]16 (objects 19, 6, 7 and 11 in Table \ref{clusids.tab},
respectively) are characterised by a present-day MF that differs
significantly from a standard Salpeter-type (I)MF; evolved to an age
of 12 Gyr using a Salpeter IMF, their luminosities will fade to well
beyond the $3\sigma_{\rm scatter}$ envelope. This conclusion remains
valid even in view of the large photometric uncertainty associated
with [WS95]355 (see Section \ref{antunc.sec}). Antennae YMCs [W99]1
and [M03] (objects 20 and 23, respectively; note the large photometric
uncertainty associated with [M03]), on the other hand, appear to have
an (I)MF that is closer to the Salpeter function down to low stellar
masses, if we assume that when the current generation of YMCs in the
local Universe evolves to GC-type ages, they should also occupy the GC
relationship. Depending on the uncertainties in the luminosity
evolution (see Section \ref{uncertainties.sec}), cluster [W99]2's
(object 15) evolved location in the $L_V-\sigma_0$ plane is also
consistent with such a Salpeter-type (I)MF. We note, however, that all
of these objects may well have non-Salpeter-type MFs, considering that
our simple modelling lets them evolve to significantly fainter
magnitudes than expected if they were to obey the well-defined Local
Group GC relationship at similar age.

In order for a YMC to survive to old age, it needs to have sufficient
low-mass stars to remain bound for a Hubble time. This condition is
met for Salpeter-type IMFs extending down to masses on the order of
0.1 M$_\odot$, but not for objects with much shallower slopes, or
(obviously) a low-mass cut-off.

Thus, from a detailed comparison between our results and those
presented in figures 6 and 7 of Mengel et al. (2002) and in Mengel
(2003), we conclude that, on average, we obtain similar predictions
for the future evolution of the Antennae YMCs, although our detailed
conclusions may differ for some of the individual objects. For
instance, while Mengel et al. (2002) suggest that [WS95]355 and
[W99]15 may be better represented by a slightly steeper than Salpeter
slope, $\alpha = 2.5$ for the full mass range from 0.1 to 100
M$_\odot$, we do not believe that the uncertainties inherent to the
data warrant such a fine distinction. While for objects [W99]1 and 2
they obtain somewhat conflicting results from their $V$ and $K$-band
data, our conclusions (based on the $V$-band data) agree for [W99]1,
but differ for [W99]2. These discrepant results may in part be
explained by the difficulty of obtaining clean cluster photometry from
ground-based ($K$) versus {\sl HST}-based ($V$) data; the difference
in M/L ratios in Mengel et al.'s (2002) between the $V$ and the $K$
band is as expected if source confusion played a more important role
in the ground-based images. In addition, in the presence of
significant mass segregation, one would also expect to obtain
different results between the $V$ and $K$-band M/L ratios (e.g.,
McCrady et al. 2003, 2005), in a similar sense as seen here. However,
the data of Mengel et al. (2002) show a general offset between the $V$
and the $K$ band for all of their objects, so that this cannot be the
only explanation. In essence, this shows the extent to which one can
rely on any individual approach; it shows, in particular, that
conclusions on the evolution of the objects that are predicted to
evolve to the area close to the 2--$3\sigma_{\rm scatter}$ transition
region in Fig. \ref{diagnostic1.fig} should be treated with caution.

Finally, most of the objects that we predict to overshoot the
$3\sigma_{\rm scatter}$ boundary by a significant amount by the time
they reach an age of 12 Gyr are located in the higher-density regions
of the system. It is likely that the ambient pressure in the
interaction region is significantly higher, and externally driven
dynamical evolution proceeds faster than in the more quiescent spiral
arm regions (Section \ref{implications.sec}); this may render invalid
the assumption that these clusters are in virial equilibrium, in
particular in view of their very young ages, of $6.6 - 10$ Myr (Mengel
et al. 2002, Mengel 2003; see Table \ref{clusids.tab}).

\subsubsection{NGC 1487 clusters}

Based on the $M/L_K$ determinations in Mengel (2003) and their
location in the (Age vs. $M/L_K$) diagram, the luminosities of YMCs
NGC 1487-1 and 2 are consistent with Salpeter-type IMF slopes down to
masses of $\sim 0.1$ M$_\odot$. Cluster NGC 1487-3, on the other hand,
has a much lower $K$-band M/L ratio for approximately the same age
(see Mengel 2003), which is indicative of a steeper IMF slope. 

Evolved to a common age of 12 Gyr in Fig. \ref{diagnostic1.fig},
clusters NGC 1487-1 and 2 are found in the boundary region between GC
stability and GC dissolution, i.e., between the 2 and 3 $\sigma_{\rm
scatter}$ envelopes. The uncertainties in the $V$-band photometry that
we obtained from our $K$-to-$V$ conversions, and also the luminosity
evolution may reduce the lengths of their luminosity evolution arrows
(see Section \ref{uncertainties.sec}), so that these objects may
potentially evolve into GC-type objects over a Hubble time (but see
Section \ref{implications.sec}).

Compared to NGC 1487-1 and 2, object NGC 1487-3, appears to be an
outlier, which may evolve to well beyond the $3\sigma_{\rm scatter}$
envelope if its present-day MF is Salpeter-like. However, we note that
the large photometric uncertainty introduced by our passband
conversion only allows us to conclude this tentatively. 

If we compare the loci of the NGC 1487 YMCs in the (Age vs. $M/L_K$)
diagram of Mengel (2003) with their expected evolution in the
$L_V-\sigma_0$ diagram of Fig. \ref{diagnostic1.fig}, we conclude that
our results are consistent with those of Mengel (2003). Clusters 1 and
2 are (perhaps marginally) consistent with Salpeter-type MFs, while
YMC 3 is characterised by an overabundance of low-mass stars compared
to clusters 1 and 2 (and compared to the standard Salpeter IMF), and
is better represented by an IMF with a steeper-than-Salpeter slope
($\alpha \approx 3$) for a stellar mass range from 0.1 to 100
M$_\odot$.

Once again, these objects are among the youngest in our sample, and as
such they may not yet be entirely virialised.

\subsubsection{M82 clusters}
\label{m82clus.sec}

When we evolve the luminosities of clusters F, MGG-9 and MGG-11 to a
common age of 12 Gyr, they are all found within $1\sigma_{\rm
scatter}$ about the GC relationship. This implies, again adopting the
assumption that all old GCs are confined to a narrow distribution in
$L_V-\sigma_0$ space and characterised by a Salpeter IMF, that these
three M82 clusters may potentially evolve into GC-type objects.
McCrady et al. (2003, 2005) suggest that all three clusters are
affected by significant mass segregation, whether primordial or
dynamical: every single YMC studied in sufficient (spatially resolved)
detail to date is known to show significant mass segregation, from the
youngest ages (see de Grijs et al. 2002a,b for a discussion). In the
presence of significant mass segregation, the estimated YMC masses are
lower limits.

McCrady et al. (2003) concluded that MGG-9 and MGG-11 are consistent
with Salpeter-like IMFs, {\it in the presence of significant
(primordial) mass segregation.} Neglecting the effects of mass
segregation, MGG-11 appears to be high-mass dominated. This scenario
seems to be confirmed by our results based on Fig.
\ref{diagnostic1.fig}. Smith \& Gallagher (2001), on the other hand,
concluded that M82-F will likely dissolve within the next $\sim 1$
Gyr. They concluded that its IMF was likely truncated at a lower mass
of 2--3 M$_\odot$, thus retaining too few low-mass stars to produce a
bound cluster over time-scales longer than a Gyr. However, McCrady et
al. (2003, 2005) provide evidence for mass segregation in cluster F
(resulting in more compact profiles at redder wavelengths), while they
also redetermine the age to be toward the lower limit of the
uncertainty range quoted by Smith \& Gallagher (2001). The latter
authors' result is also affected by a somewhat uncertain correction
for the saturated cluster centre in the {\sl HST} $V$-band image.
Taking all of these effects together, McCrady et al. (2003, 2005)
conclude that M82-F may be deficient in low-mass stars (i.e., a simple
application of SSP models to the observed M/L ratio suggests a
low-mass cut-off at $\sim 2$ M$_\odot$), although in view of the
significant mass segregation present, it is equally likely
characterised by a ``standard'' IMF. These results support our
conclusion.

\subsubsection{M83 (NGC 5236) clusters}

Of the two M83 clusters in our sample, object NGC 5236-502 appears to
be characterised by a standard Salpeter IMF, based on the fact that
adopting this IMF will let the YMC evolve to a location close to the
old GC relationship. This is fully consistent with the conclusion
reached by Larsen \& Richtler (2004), based on their more complex
analysis of the cluster's dynamical mass and its corresponding M/L
ratio. Cluster NGC 5236-805, however, appears to overshoot the
$2\sigma_{\rm scatter}$ envelope somewhat, if it were governed by a
similar initial and/or present-day MF, although the uncertainties
inherent in the luminosity evolution (see Section
\ref{uncertainties.sec}) still allow for this object to have a
close-to-Salpeter MF. Thus, we conclude that our results for this
object are also consistent with Larsen \& Richtler's (2004)
independent assessment.

\subsubsection{NGC 1569-A1}

The measurements for NGC 1569-A1 are affected by significant
uncertainties. The original high-dispersion spectra of Ho \&
Filippenko (1996a) are contaminated by flux from the its binary
companion cluster, A2, which was first realised by De Marchi et
al. (1997). However, since A1 is almost twice as bright as A2, De
Marchi et al. (1997) argued that the basic velocity dispersion
measurement of Ho \& Filippenko (1996a) still reflects that of the
main component, A1. In addition, because of the contamination by A2,
the age determination of component A1 is affected by significant
uncertainties (see Table \ref{clusids.tab}). For the purpose of the
present paper, we have used the most up-to-date photometry of De
Marchi et al. (1997) and the best age determination of $\sim 8$ Myr
(Hunter et al. 2000, Origlia et al. 2001). When we evolve the
cluster's luminosity to an age of 12 Gyr, it is found on the
$2\sigma_{\rm scatter}$ envelope of the GC relation. The uncertainties
inherent in the luminosity evolution are such that any correction will
result in this evolution being reduced and thus the cluster would end
up closer to the GC relation. Therefore, we predict that NGC 1569-A1
will likely become an old GC (in the absence of external disruptive
forces; see Section \ref{implications.sec}). As a consequence, we also
suggest that the cluster's IMF may be close to the standard Salpeter
IMF. Our conclusions are consistent with those of De Marchi et
al. (1997), based on their analysis of the evolution of the M/L ratio,
assuming a Salpeter IMF down to the hydrogen-burning limit, and with
Origlia et al. (2001), based on SSP fits governed by variety of
IMFs. Our results are also consistent with Ho \& Filippenko (1996a),
despite different assumptions used for the mass determinations; these
authors also concluded that -- to a first approximation -- the NGC
1569-A IMF appeared to be similar to that of typical Galactic GCs.

\subsubsection{NGC 1705-I}

Ho \& Filippenko (1996b) concluded, using a similar approach as for
NGC 1569-A (i.e., A1 and A2 combined), that NGC 1705-I has all the
properties (M/L ratio, radius, mass) of a young, metal-rich GC (but
note the caveat mentioned above regarding their mass determinations).
In the most recent detailed study of the stellar content of NGC
1705-I, V\'azquez et al. (2004) conclude -- based on {\sl HST}/STIS
spectroscopy and an analysis of the cluster's M/L ratio -- that there
is no significant evidence for an anomalous IMF at the low-mass end,
contrary to previous suggestions (see references in V\'azquez et
al. 2004). This is fully consistent with the location of the YMC in
our diagnostic $L_V-\sigma_0$ diagram when evolved to an age of 12
Gyr.

\subsubsection{Clusters in NGC 4214 and NGC 4449}

Larsen et al. (2004) obtained high-dispersion spectra for four YMCs in
the dwarf irregular galaxies NGC 4214 and NGC 4449. For all clusters,
they find M/L ratios that are similar to or slightly higher than for a
Salpeter or Kroupa01-type IMF. They thus rule out any present-day MF
that is deficient in low-mass stars compared to these IMFs. They
conclude that these objects might therefore evolve to become old GCs
over a Hubble time. This conclusion is fully supported by the location
of the evolved YMCs in our diagnostic diagram of
Fig. \ref{diagnostic1.fig}.

\subsubsection{NGC 6946-1447}

Just as for the YMCs in NGC 4214 and NGC 4449, Larsen et al. (2004)
also conclude that the present-day MF of NGC 6946-1447 resembles a
Salpeter or Kroupa-type MF quite closely. They essentially confirmed
their earlier result for this cluster (Larsen et al. 2001) where they
concluded that the estimates for its dynamical mass and its
photometric mass based on SSPs governed by a Salpeter IMF were similar
within the model uncertainties. Thus, this object also has the
potential of evolving into an old GC if not disrupted prematurely by
external factors. This is again fully consistent with the cluster's
evolved location in our diagnostic $L_V-\sigma_0$ diagram.

\subsubsection{NGC 7252-W3}

Finally, in a detailed spectroscopic and photometric study, Maraston
et al. (2004) conclude that the dynamical virial mass for NGC 7252-W3,
based on their newly obtained high-dispersion spectroscopy, is in
excellent agreement with photometric values previously estimated
(Schweizer \& Seitzer 1998, Maraston et al. 2001) from the cluster
luminosity by means of stellar M/L ratios predicted by SSP models with
a Salpeter IMF down to stellar masses of $\sim 0.1$ M$_\odot$. While
this conclusion is consistent, within the uncertainties, with the
object's evolved location in our diagnostic diagram of
Fig. \ref{diagnostic1.fig}, its velocity dispersion places it in the
realm of the nuclear clusters and UCDs, so that caution needs to be
exercised when comparing results in this context.

\subsubsection{Concluding remarks}

Thus, it appears that the simple diagnostic $L_V-\sigma_0$ diagram
results in consistent predictions regarding the evolution of YMCs in
the local Universe, without the need to convert the observed velocity
dispersions into dynamical masses and thus introducing additional
assumptions and their associated uncertainties. Discrepancies between
predictions on the YMCs' evolutionary fate resulting from the
application of different methods serve as a useful diagnostic
providing insight into the likely range of uncertainties involved in
any of these predictions. We note that our predictions should be
treated as first-order predictions (as should those resulting from
using dynamical mass estimates). They do not include external factors
that might speed up the dissolution of otherwise firmly bound star
clusters; we will address this issue in Section
\ref{implications.sec}. Nevertheless, to first order, the fact that
most clusters, when evolved using a standard solar-neighbourhood
Salpeter-type IMF, appear to end up close to the GC relationship
(although systematically somewhat to fainter magnitudes) instills some
confidence in the universality of this IMF for extragalactic YMCs,
leaving little leeway for significant IMF variations, {\it assuming
that they may potentially survive for a Hubble time}. We note in
passing that dynamical evolution of $\sigma_0$ will tend to move our
sample clusters even closer to the old GC relation, adding weight to
this conclusion (see Section \ref{sigma.sec}). We will discuss those
objects that still appear to overshoot the GC relation in more detail
in Section \ref{implications.sec}.

Finally, in Fig. \ref{diagnostic1.fig} we have also included the
relevant data points for the compact LMC and SMC clusters younger than
10 Gyr at the present time (open squares; Dubath et al. 1993, 1997;
photometry from Bica et al. 1996, de Freitas Pacheco, Barbuy \& Idiart
1998). If these objects are characterised by a Salpeter-type
present-day MF and IMF, as is supported by observational evidence
(see, e.g., de Grijs et al. 2002a,b for a representative sample of
compact LMC clusters), they will fade by up to $\sim 4$ mag (and in
most cases by more than $\sim 1.5$ mag) before they reach an age of 12
Gyr. However, very few of the compact LMC and SMC clusters extend to
fainter absolute magnitudes than contained within the $2\sigma_{\rm
scatter}$ envelope of the best-fitting GC relation. This implies
either that cluster disruption, at least in the Magellanic Clouds,
must occur before a cluster fades to this limit, or that the old GC
relation for the lower-density LMC environment is significantly
different from (and much broader than) that in the Galaxy and M31. If
we assume that the GC relation is independent of environment, as seems
to be suggested by the good agreement of the old GCs in the Local
Group, we predict that at least half of the LMC and SMC clusters
younger than 10 Gyr will dissolve before reaching GC-type ages. The
small number of LMC and SMC clusters currently beyond the $2
\sigma_{\rm scatter}$ boundary may either be caused by statistical
sampling effects or perhaps we have caught objects in the process of
dissolution. Once again, the presence of these objects gives a good
indication of the uncertainties involved in using the $L_V-\sigma_0$
diagnostic diagram: there is most likely a transition region in the
diagram where clusters may or may not evolve to, depending on the
details of their internal and environmental properties. In this
context, we note that the LMC provides a fairly low-density stellar
environment, particularly outside the central, barred region.

The two Magellanic Cloud objects toward brighter magnitudes than the
best-fitting GC relationship are the youngest LMC cluster for which we
have velocity dispersion information, NGC 1818 (25 Myr; de Grijs et
al. 2002a) and NGC 419 in the SMC. If they are characterised by
Salpeter-type IMFs down to $\sim 0.1$ M$_\odot$ (cf. de Grijs et
al. 2002b), these objects are likely to fade by $\sim 5$ and $\sim 2$
mag, respectively. Judging from their location in Fig.
\ref{diagnostic1.fig}, we predict that while NGC 419 may possibly
become an object equivalent to NGC 121 (the only GC-equivalent object
in the SMC), NGC 1818 will likely disperse long before. We emphasise
that in this case we have independent measurements of the cluster's
present-day MF (de Grijs et al. 2002a,b), so that this is a firm
conclusion.

In this context, it is interesting to compare these results for the
massive, compact star clusters in the Local Group to the Galactic open
clusters. The Galactic cluster population exhibits a clear dichotomy,
in the sense that all Galactic GCs are older than $\sim 10$ Gyr, while
few Galactic open clusters are older than a few Gyr. If we include the
roughly 40 Galactic open clusters with relevant observational data
(Lohmann 1972, Sagar \& Bhatt 1989) in our diagnostic diagram, they
occupy a well-delineated region centred at $\log( \sigma_0/\mbox{km
s}^{-1}) \sim -0.25$, and lying on the extrapolation of the GC
relationship. Considering that, if they were governed by a
Salpeter-type IMF down to the hydrogen-burning limit, they would fade
by at least another 2 mag, their location in the $L_V-\sigma_0$
diagram is consistent with the observational fact that there are no
known open clusters of typical GC age in the Galaxy.

\section{Assessment of the uncertainties}
\label{uncertainties.sec}

Having established that, to first order, the $L_V-\sigma_0$ diagram
provides us with a diagnostic tool to assess the similarities (and
differences) of YMCs compared to old GCs, we will now assess the
uncertainties inherent to this approach. In Section \ref{mv.sec} we
will first address the uncertainties related to the evolution in
luminosity of a given cluster. Subsequently, in Section
\ref{sigma.sec} we will present the results of detailed {\it N}-body
simulations to obtain a feeling for the uncertainties associated with
the evolution of the central velocity dispersion over a Hubble time.

\subsection{Luminosity evolution}
\label{mv.sec}

The main issue we need to address regarding the luminosity evolution
of our sample YMCs, as represented by the ``luminosity evolution
arrows'' in Fig. \ref{diagnostic1.fig}, is the accuracy of the arrow
lengths. In addition, we will address a number of issues related to
the accuracy of the photometric measurements of the objects
themselves. Regarding the former, the key issues to be discussed are
the dependence of the luminosity evolution on (i) metallicity and (ii)
the adopted IMF (and, therefore, on the adopted SSP models).

\begin{figure}
\psfig{figure=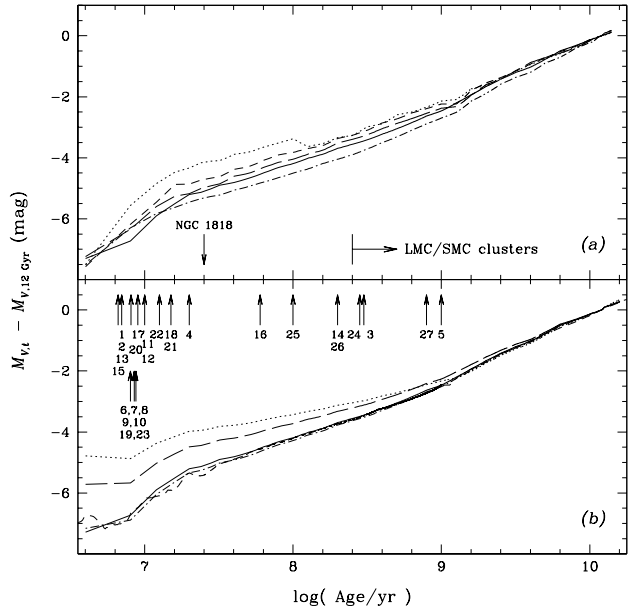,width=8.7cm}
\caption[]{Uncertainty assessments in the luminosity evolution of the
YMCs identified in Table \ref{clusids.tab}. As a function of their
age, we display the uncertainties in the lengths of the dotted arrows
in Fig. \ref{diagnostic1.fig} caused by {\it (a)} metallicity
variations, for a Salpeter IMF covering a mass range from 0.1 to 100
M$_\odot$ and adopting the GALEV SSP models, and {\it (b)} variations
in the IMF, for solar metallicity. The sample clusters are identified
at their appropriate ages. The line styles in panel {\it (a)}
correspond to metallicities of $0.02 Z_\odot$ (dotted), $0.2 Z_\odot$
(short dashed), $0.4 Z_\odot$ (long dashed), $Z_\odot$ (solid), and
$2.5 Z_\odot$ (dot dashed). In panel {\it (b)}, they refer to a
Salpeter IMF for GALEV and Starburst99 SSPs (solid and short dashed,
respectively), and GALEV SSPs computed for a Scalo (dotted), Kroupa01
(dot dashed) and KTG93 (long dashed) IMF. The mass range covered is
from 0.1 to 100 M$_\odot$ for all GALEV SSPs, while the Starburst99
SSPs are truncated at low mass and cover masses from 1 to 100
M$_\odot$.}
\label{uncertainties.fig}
\end{figure}

In Fig. \ref{uncertainties.fig}a, we show the expected length of the
luminosity evolution arrow as a function of cluster age,
($M_{V,t}-M_{V,12\;{\rm Gyr}}$), for the five different metallicities
included in the GALEV SSPs. For the purposes of this discussion, we
have adopted a Salpeter IMF, covering stellar masses from 0.1 to 100
M$_\odot$. It is clear that the effect of adopting an incorrect
metallicity is roughly constant as a function of age, and amounts to
an error of $\lesssim 0.8$ mag over the entire age range spanned by
our YMC sample if solar metallicity were incorrectly assumed. The
effect decreases slightly for cluster ages $\gtrsim 10^9$ yr. We note
that we have taken great care to adopt the most appropriate
metallicity for our sample YMCs (see Table \ref{clusids.tab}), so that
we are confident that we have minimised the uncertainties associated
with the choice of cluster metallicity.

\begin{figure}
\psfig{figure=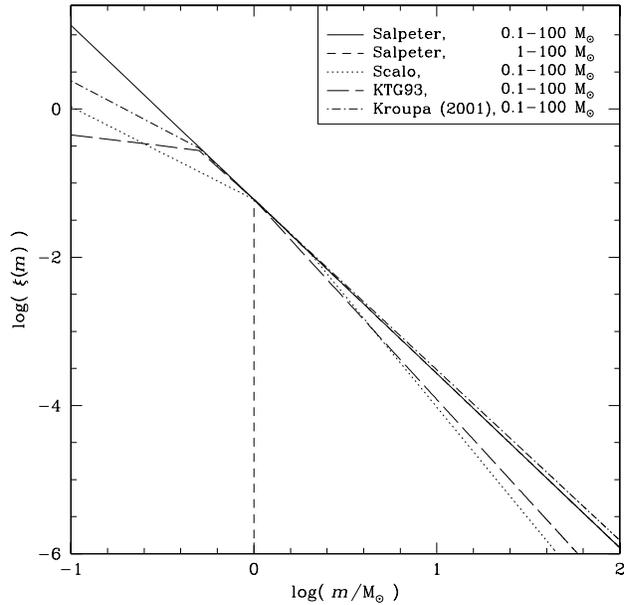,width=8.7cm}
\caption[]{Representations of the IMFs used in this paper, clearly
showing the relative importance of the contributions of the low
vs. high-mass stars. All IMFs have been normalised to reproduce the
standard Salpeter IMF at 1 M$_\odot$, while the standard Salpeter IMF
has been normalised to contain a total mass of 1 M$_\odot$. The
different line styles refer to a Salpeter IMF for GALEV and
Starburst99 SSPs (solid and short dashed, respectively), and GALEV
SSPs computed for Scalo (dotted), Kroupa01 (dot dashed) and KTG93
(long dashed) IMFs. The mass range covered is from 0.1 to 100
M$_\odot$ for all GALEV SSPs, while the Starburst99 SSPs are truncated
at low masses and cover masses from 1 to 100 M$_\odot$.}
\label{imfs.fig}
\end{figure}

Secondly, we explore the effects of varying the IMF, $\xi(m) \propto
m^\alpha$. We consider the effects of varying both the slope,
$\alpha$, and the low-mass cut-off of the IMF. In order to do so, we
calculated the age dependence of the length of the ``evolution
arrows'' in Fig. \ref{uncertainties.fig}b for five different IMF
representations, and solar metallicity. Except for IMF (ii) below,
where we use the Starburst99 SSPs (Leitherer et al. 1999), we use the
GALEV SSPs in all cases, and assume the IMF to cover the mass range
from 0.1 to 100 M$_\odot$. We consider the following IMFs, the effects
of which on the luminosity evolution are shown in
Fig. \ref{uncertainties.fig}b:
\begin{enumerate}

\item the ``standard'' Salpeter IMF, for masses between 0.1 and 100
M$_\odot$, and $\alpha = -2.35$ for the entire mass range;

\item the $\alpha = -2.35$ Salpeter IMF, but for the mass range $1-100$
M$_\odot$;

\item the Scalo (1986) IMF, for masses $0.1 < m/{\rm M}_\odot < 100$,
characterised by
\[ \alpha = \left\{ \begin{array}
{r@{\quad;\quad}l}
-1.25 & m < 1 {\rm M}_\odot \\
-2.35 & 1 < m/{\rm M}_\odot < 2 \\
-3.00 & m > 2 {\rm M}_\odot
\end{array} \right.  \]

\item the KTG93 IMF, with
\[ \alpha = \left\{ \begin{array}
{r@{\quad;\quad}l}
-0.3 & m \le  0.5 {\rm M}_\odot \\
-2.2 & 0.5 \le m/{\rm M}_\odot \le 1.0 \\
-2.7 & m > 1.0 {\rm M}_\odot
\end{array} \right.  \]
and

\item the Kroupa01 IMF:
\[ \alpha = \left\{ \begin{array}
{r@{\quad;\quad}l}
-1.3 & m < 0.5 {\it M}_\odot \\
-2.3 & m > 0.5 {\it M}_\odot
\end{array} \right.  \]
\end{enumerate}

Figure \ref{imfs.fig} displays the functional forms of these IMFs,
normalised to a standard Salpeter IMF at 1 M$_\odot$, which contains a
total mass of 1 M$_\odot$. This standard Salpeter IMF is shown as the
solid line, and is used as reference in the following. Except for the
truncated Salpeter IMF (short-dashed line), the other, more realistic
IMFs are characterised by a turnover at or below 1 M$_\odot$, and an
enhanced contribution of intermediate-mass stars ($\sim 1-10
\mbox{M}_\odot$) compared to the full Salpeter IMF. In all cases,
however, they are dominated by the lower-mass stars and provide,
therefore, a solid basis for any compact virialised system to survive
for up to a Hubble time (e.g., Gnedin \& Ostriker 1997, Goodwin 1997a,
Smith \& Gallagher 2001, Mengel et al. 2002).

It is clear that the effects on the luminosity evolution arrow of
varying the IMF are significant for all ages below several $\times
10^8 - 10^9$ yr.  Any correction to the length of the luminosity
evolution arrow caused by a significant change in the IMF (for the
IMFs discussed in this paper) is in the sense that the length of the
arrow will be {\it reduced}; for the Kroupa01 and truncated Salpeter
IMFs the effect is expected to be negligible. Thus, by adopting a more
realistic IMF than the standard Salpeter representation (such as the
KTG93 IMF, which accurately describes the solar neighbourhood IMF),
those clusters that in our current diagnostic diagram of
Fig. \ref{diagnostic1.fig} would evolve to locations well beyond the
$3\sigma_{\rm scatter}$ envelope of the GC relationship if they were
characterised by a Salpeter-type IMF down to the hydrogen-burning
limit might well evolve to a location within $\sim 2\sigma_{\rm
scatter}$. In addition, if we had assumed any more realistic IMF
description for the luminosity evolution of our sample YMCs, the
evolved loci of most of these objects might have scattered more
symmetrically around the best-fitting GC relation, instead of
systematically ending up on the faint side of the correlation (we have
confirmed this for the case of the KTG93 IMF).

Based on the currently available data, we cannot draw firm conclusions
on the actual (I)MFs of our sample clusters. Detailed follow-up {\it
N}-body simulations, including the effects of primordial and dynamical
mass segregation, and of varying binary fractions, are required to
address this issue more robustly. This is, however, beyond the scope
of the present work. On the other hand, the fact that most clusters,
when evolved using a standard solar-neighbourhood Salpeter-type IMF,
appear to end up close to the GC relationship is suggestive of the
near-universality of an IMF for extragalactic YMCs of any of the
currently fashionable forms discussed in this paper. Based on the
available evidence, it is therefore more likely that the six YMCs that
appear to have a central velocity dispersion that is significantly too
large for their mass (luminosity) will disperse before reaching
GC-type ages, than that they were characterised by significantly
different {\it initial} MFs (and possibly very different {\it
present-day} MFs; see also Section \ref{implications.sec}).

Thirdly, there are a number of observational uncertainties that affect
the accuracy of the location of the data points at the present
epoch. Some of the sample YMCs are affected by significant extinction
in their host galaxies, so that any extinction correction introduces
uncertainties in the clusters' location at the present
time. The objects most affected by these uncertainties are
\begin{itemize}
\item {\bf NGC 6946-1447}: $A_{V,{\rm Gal}} = 1.13$ mag (Schlegel et
al. 1998);

\item {\bf NGC 1042-NC}: based on $I$-band photometry, only corrected
for Galactic extinction. We believe that the main uncertainty in the
photometry of this cluster is related to our assumption of it being a
clean SSP, as discussed above;

\item {\bf IC 342-NC}: $A_V = 2.5$ mag (McCall 1989, Madore \&
Freedman 1992), but patchy and variable. B\"oker et al. (1999)
measured $A_K \sim 0.45$ mag toward the YMC, equivalent to $A_V \sim
4.0$ mag, with an uncertainy of $\Delta A_V \sim 0.9$ mag due to the
patchiness of the extinction;

\item {\bf NGC 1614-NC1,2}: based on bolometric luminosities, derived
from mid-infrared observations, so that the accuracy of the conversion
depends on the accuracy of the bolometric correction adopted. In
addition, $A_V \sim 4.7$ mag, in a clumpy distribution;

\item {\bf M82-F}: $E(B-V) = 0.9 \pm 0.1$ mag (Smith \& Gallagher
2001, but see McCrady et al. 2003). McCrady et al. (2005) conclude
that their $H$-band spectra are negligibly affected by extinction,
while $A_{\rm F814W} = 0.5 \pm 0.2$ mag;

\item {\bf M82 MGG-9} and {\bf MGG-11}: photometry based on
near-infrared {\sl HST} observations; $A_{\rm F160W} = 2.1 \pm 0.5$
and $1.4 \pm 0.5$ mag, respectively (McCrady et al. 2003). Translated
to the $V$ band, the extinction becomes considerable, at $A_V \sim 12
\pm 3$ and $8 \pm 3$ mag, respectively;

\item {\bf NGC 5236 clusters}: $A_{B,{\rm Gal}} = 0.284$ mag (Schlegel
et al. 1998). The internal extinction $A_B = 1.0 \pm 0.2$ mag, and
$1.0 \pm 0.5$ mag for NGC 5236-502 and NGC 5236-805, respectively
(Larsen \& Richtler 2004). Cluster 502 is located close to both a
conspicuous dust lane, and to a fainter, bluer companion cluster; both
objects are unresolved at ground-based spatial resolution.

\item {\bf NGC 4214-13}: $A_B = 1.09 \pm 0.05$ mag (Larsen et
al. 2004);

\item the {\bf Antennae clusters [WS95]355} (photometry based on $I$
band data, since only an upper limit could be obtained in $V$;
significant extinction), {\bf [WS95]331} and {\bf [M03]} (both based
on $K$-band photometry; significant extinction). Based on a comparison
of Mengel et al. (2001, 2002), Antennae YMCs {\bf [W99]1, 10}, and
{\bf 16} are affected by $A_V = 0.6 \pm 0.3$, $0.3 \pm 0.3$ and $0.3
\pm 0.3$ mag of extinction; the other Antennae objects are more
highly extincted, although the details are lacking in the original
papers.

\item the {\bf NGC 1487 clusters}: based on ground-based $K$-band
photometry; no extinction estimates available.
\end{itemize}

\begin{table*}
\caption[ ]{\label{tab:nbody_props} Parameters of the simulated
$N$-body clusters shown in
Fig.~\protect\ref{fig:nbody_evol}. Respectively, the columns show the
model name, initial mass $M_{\rm i}$ (in solar masses), metallicity
$Z$, initial hard binary fraction $f_{\rm b}$ (percentage), final mass
$M_{\rm f}$ (in solar masses), the length of the simulation $t_{\rm
end}$ (in Gyr), the external potential, the type of cluster orbit
(circular or eccentric) in the external potential and the source of
the $N$-body data. The external potentials used were (1) PT = point
mass; (2) MW = linearised Milky Way disc potential.}

\begin{center}
\begin{tabular}{rllrcrccc}
\hline
\hline
\multicolumn{1}{c}{Name} & \multicolumn{1}{c}{$M_{\rm i}$} &
\multicolumn{1}{c}{$Z$} & \multicolumn{1}{c}{$f_{\rm b}$} & 
\multicolumn{1}{c}{$M_{\rm f}$} &  \multicolumn{1}{c}{$t_{\rm end}$} & \multicolumn{1}{c}{Ext. Potential} & 
\multicolumn{1}{c}{Orbit} & \multicolumn{1}{c}{Ref.$^a$} \\
\hline

Circ1 & $2.4\times 10^3$ & 0.02 & 0 & 550 & 2.1 & PT & Circle & 1\\
Circ2 & $2.3\times 10^3$ & 0.02 & 50 & 405 & 1.6 & PT & Circle & 1\\
Ecc1 & $2.4\times 10^3$ & 0.02 & 0 & 130 & 1.3 & PT & Eccentric & 1\\ \hline

Circ3 & $4.9\times 10^4$ & 0.001 & 0 & $1.6\times 10^4$ & 12 & MW & Circle & 2\\
Circ4 & $5.2\times 10^4$ & 0.001 & 5 & $1.5\times 10^4$ & 12 & MW & Circle & 2\\
Circ5 & $4.9\times 10^4$ & 0.0002 & 0 & 770 & 9 & PT & Circle & 2\\
\hline
\end{tabular}
\end{center}
$^a$References: 1, Wilkinson et al. (2003); 2, Hurley et al. (2005)
\end{table*}

However, while these uncertainties are significant, the respective
authors in the original papers have taken great care to correct for
these effects as well as possible, while we have applied additional
corrections where it was deemed necessary.

Finally, we need to be aware of the potential effects caused by
stochasticity in the IMF. At masses of up to a few $\times 10^4$
M$_\odot$, IMF sampling effects become noticeable and significant
(Lan\c{c}on \& Mouhcine 2000, Bruzual 2002, Bruzual \& Charlot
2003). The increase in the scatter around the GC relationship toward
lower central velocity dispersions may be due to the effects of poor
IMF sampling -- some of this scatter may also be due to the increased
importance of external perturbations for low-mass clusters. However,
for our extragalactic YMCs, these effects are likely minimally, if at
all, important. Because of the current technical limitations, we can
only obtain high-dispersion spectroscopy of the highest mass YMCs,
which are expected to have well-sampled IMFs.

In summary, the most important internal factor affecting the accuracy
of the luminosity evolution of our sample YMCs is related to the
functional form of the IMF assumed when applying the evolutionary
corrections. However, based on the apparent universality of the IMF in
a wide variety of environments (see, e.g., the review by Gilmore
2001), using a single IMF description for the entire sample seems
reasonable\footnote{We also note that the maximum differences in
luminosity evolution of the IMFs presented in Fig.
\ref{uncertainties.fig}b from the youngest YMC age observed, at $\sim
6$ Myr, to 12 Gyr is $\lesssim 2.2$ mag. This is well within the
uncertainties allowed for by using the $2 \sigma_{\rm scatter}$
boundary as our diagnostic, so that the use of a single IMF
description for the full YMC sample seems justified.} and has the
potential to provide valuable and robust insights into the future fate
of a given sample of YMCs.

\subsection{Dynamical evolution}
\label{sigma.sec}

In the previous section, we quantified the uncertainties in the
estimated luminosities our clusters would have at a fiducial, common
age of 12 Gyr. Our estimates implicitly assumed that the tracks
followed by clusters in the $L_V - \sigma_0$ plane are determined only
by stellar evolution, and thus neglected the role of dynamical
evolution. In particular, the central velocity dispersion of the
cluster was assumed to remain constant throughout the evolution. In
fact, there are a number of competing factors that affect the
evolution of the cluster velocity distribution. For example, mass loss
from stellar evolution or due to tidal stripping by an external tidal
field may reduce the overall velocity dispersion of the cluster, while
the long-term evolution toward core collapse will tend to produce an
increase in velocity dispersion in the central parts of the
cluster. In addition, mass segregation of the more luminous (i.e.,
more massive) stars could potentially give rise to a fall in the
measured central cluster dispersion as these stars will dominate the
cluster light, and hence their smaller velocities will serve to reduce
the observed dispersion in the core regions. On the other hand, mass
segregation of binaries will tend to inflate the measured central
velocity dispersion, as the orbital velocities will contribute to the
observed cluster dispersion. $N$-body simulations of star clusters are
the most reliable way to study the combined effects of stellar and
dynamical evolution (both internal and external) on cluster properties
(e.g., Portegies Zwart 2002, Baumgardt \& Makino 2003, Wilkinson et
al. 2003, Dehnen et al. 2004). In this section, therefore, we present
results from $N$-body simulations, in order to quantify the likely
evolution of a cluster in the $L_V - \sigma_0$ plane.

The $N$-body clusters presented in this section comprise two separate
sets of models: (i) low-mass clusters from Wilkinson et al. (2003),
with masses of about $2400$ M$_\odot$; (ii) intermediate-mass clusters
with masses of about $5\times 10^4$ M$_\odot$ from J. Hurley
(priv. comm. and Hurley et al. 2005). All simulations were performed
using the {\sc nbody4} code (Aarseth 1999) running on the GRAPE-6
special purpose computer boards (Makino 1997) at the Institute of
Astronomy, Cambridge and the American Museum of Natural History, New
York. {\sc nbody4} is a direct $N$-body code, which incorporates
stellar evolution routines based on parameterised functions (Hurley et
al. 2001) to follow the evolution, on a star-by-star basis, of the
single stars and binaries in the cluster. Given the realistic nature
of the simulations, it is possible to analyse the model clusters in
precisely the same way as observed clusters. The importance of
studying simulated cluster evolution in terms of directly observable
quantities has been emphasised by a number of authors (e.g., Wilkinson
et al. 2003, Portegies Zwart 2001).

The relevant parameters of the simulated clusters are given in
Table~\ref{tab:nbody_props}. More details can be found in Wilkinson et
al. (2003) and Hurley et al. (2005). The stellar IMF of KTG93 was used
to populate the mass spectrum of each cluster -- lower and upper mass
cut-offs of $0.1$ M$_\odot$ and $50.0$ M$_\odot$, respectively, were
assumed. The low-mass cluster simulations were carried out in the
external potential of a point mass of mass $9\times10^9$ M$_\odot$ and
ran for between 1.3 and 2.1 Gyr, by which time each cluster had lost
more than 75 per cent of its mass. Two of the intermediate-mass
clusters (models Circ3 and Circ4) were evolved in a linearised
approximation of the Milky Way disc potential at the position of the
Sun: these clusters contained approximately $25$ per cent of their
initial mass after $12$ Gyr. Model Circ5, on the other hand, was
placed on a circular orbit at a radius of $4$ kpc from a point mass of
mass $4.5\times 10^{10}$ M$_\odot$. This cluster had lost more than
$98$ per cent of its mass when the simulation was stopped at $9$
Gyr. Thus, although the model clusters are necessarily less massive
than the YMCs in the observational sample (due to computational
constraints), they nevertheless span a range of masses, binary
fractions, external potentials and orbits. Most importantly, the
sample includes both clusters that disrupt rapidly and some that
survive to late times, placing useful constraints on the expected
evolution in the $L_V - \sigma_0$ plane for clusters experiencing
widely varying degrees of external perturbation.

\begin{figure}
\psfig{figure=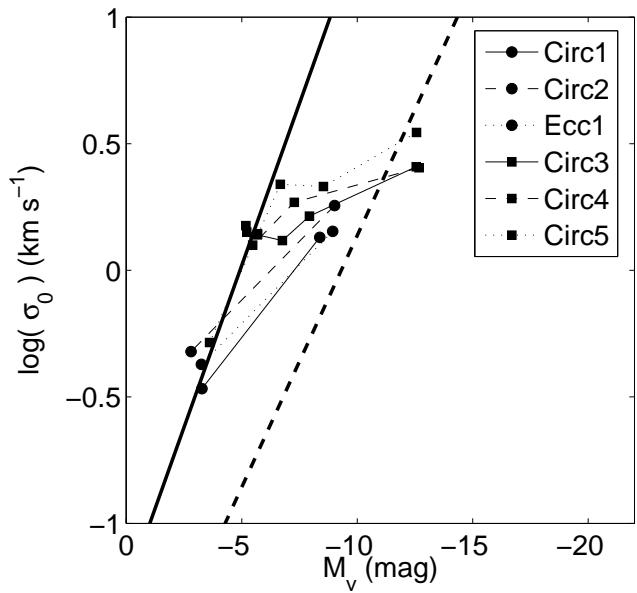,width=8.4cm}
\caption[]{Evolution of central velocity dispersion and absolute
magnitude of $N$-body clusters. The lower curves (with solid points)
denote the low-mass models: data are plotted for time $= 0$ and
$t_{\rm end}$. The upper curves (with solid squares) are for the
intermediate-mass clusters. The output times for these models are (a)
Circ3: $0,1,5,9,10,12$ Gyr (b) Circ4: $0, 2.5, 10.5, 12$ Gyr (c) Circ5:
$0, 1, 5, 9$ Gyr. For all models, the evolution proceeds from right to
left in this figure. The thick solid and dashed lines indicate the
observed relations for the Local Group GCs, and for the youngest YMCs,
respectively (Section \ref{implications.sec}).}
\label{fig:nbody_evol}
\end{figure}

In order to facilitate the comparison of the simulation results with
the observed clusters, we need to estimate the central velocity
dispersions and absolute magnitudes of the simulated clusters. The
absolute magnitudes of the clusters were calculated simply by adding
up the individual stellar luminosities of all the stars in each
cluster. In order to reduce the numerical noise in the estimate of the
velocity dispersions, the velocity dispersions were calculated for
three perpendicular lines of sight and the results averaged. For each
line of sight, the projected radius containing half the cluster light
was calculated and only those stars that lay within this radius were
included in the dispersion calculation. For binary stars, a random
orientation was chosen for the binary orbit and we assumed that all
binaries were observed at apocentre (where the stars spend most of
their time). The relative motion of the stars in the frame of their
centre-of-mass was calculated based on their masses and the semi-major
axis and ellipticity of the orbit. The full space motions of the stars
in the frame of the cluster were then calculated and the line-of-sight
component of this motion was included in the cluster dispersion
calculation.

In order to mimic the process by which a velocity dispersion is
measured from the line widths in an integrated spectrum of an observed
cluster, the distribution of line-of-sight velocities was fitted by a
Gaussian of mean velocity $\overline{v}$ and dispersion $\sigma$. For
models Circ3, Circ4 and Circ5 we constructed a luminosity-weighted
cumulative velocity distribution from the individual stellar
velocities and luminosities, and found the Gaussian distribution whose
cumulative distribution was a best fit in the least-squares
sense. This procedure ensures that bright stars contribute more to the
dispersion calculation than fainter stars, as is the case in real
observations. As Boily et al. (2005) point out, it is essential to
take account of this effect when comparing simulated and observed
clusters. For models Circ1 and Ecc1, luminosity information was not
available for the stars, and for model Circ2 the calculation produced
unacceptably noisy results due to the small numbers of stars. For
these models, therefore, stars of all masses were weighted
equally. Their velocity evolution should therefore be taken as
indicative only. In all cases, following the initial calculation of
$\overline{v}$ and $\sigma$, the estimates were refined by removing
stars whose velocities were more than $3\sigma$ away from the mean of
the sample and recalculating $\overline{v}$ and $\sigma$.  This
process was repeated until removing further outliers had a negligible
impact on the estimated dispersion. A direct calculation of the
dispersion of line-of-sight velocities would be skewed by the presence
of the highest-velocity binaries, whose orbital velocities greatly
exceed the dispersion of the cluster, and which contribute
non-Gaussian tails to the velocity distribution. Our procedure reduces
their impact on the estimated dispersion in a manner consistent with
observational techniques, which are generally insensitive to the
presence of low-level, non-Gaussian tails such as those produced by
binaries (see, e.g., Larsen et al. 2004). Generally, Gaussian fitting
is used to determine the centre and FWHM of a spectral line, thereby
ignoring any non-Gaussian tails which might indicate the presence of a
binary population.

Figure \ref{fig:nbody_evol} shows the evolution of our simulated
clusters in the $L_V - \sigma_0$ plane. For the low-mass clusters,
results are presented for times $t=0$ and $t=t_{\rm end}$. For the
intermediate-mass clusters, output times up to 12 Gyr are shown (with
the exception of model Circ5, which was disrupted after 9 Gyr). There
are several points to note from this figure. First, the evolution of
$\sigma_0$ for clusters which survive to late times (models Circ3 and
Circ4) is quite limited, particularly for model Circ4, which contains
a population of primordial binaries. For these models, the change in
central velocity dispersion is $\Delta(\log\sigma_0) < 0.3$ dex over
the course of 12 Gyr. Thus, our assumption that the evolution of the
observed clusters in Fig. \ref{diagnostic1.fig} is dominated by the
evolution of their absolute magnitudes is reasonable\footnote{The
additive effect of the evolution in $\sigma_0$ on top of the
luminosity evolution is that a few of our sample clusters deemed on
the verge of disruption based on their evolution in luminosity alone,
are now thought to evolve to more safe / stable loci inside the
$2\sigma_{\rm scatter}$ envelope.} and therefore our conclusions based
on that figure are unchanged. Figure \ref{fig:nbody_evol} also shows
that clusters that disrupt during the course of the simulations
exhibit more significant evolution of $\sigma_0$. This emphasises the
role of external factors in determining the late-time evolution of
clusters in the $L_V - \sigma_0$ plane. Thus, as we already made clear
above, not all the YMCs in our sample that have the potential to
survive to late times will necessarily do so if their external
environment is too extreme.

Secondly, the presence of large numbers of binaries can significantly
affect the observed velocity dispersion, especially at late times. As
expected, binaries tend to inflate the velocity dispersion of the
cluster, as a comparison of the evolution of models Circ2 and Circ4
(which contain primordial binaries) with that of models Circ1 and
Circ3 (which initially contain only single stars) shows. The effect
initially increases with time as mass segregation draws the binaries
to the centre of the cluster due their larger masses, leading to an
increased binary fraction in the central regions. Toward the end of
the simulations, binary-single star encounters expel sufficient
numbers of short-period binaries to move the observed dispersion
toward that of the cluster without primordial binaries.

Finally, for all simulated clusters, $\sigma_0$ decreases with time.
Interestingly, this leaves the majority of our simulated clusters very
close to the Local Group GC relation (shown as the solid line in Fig.
\ref{fig:nbody_evol}). We will return to the significance of this fact
in Section~\ref{implications.sec}, where we will suggest that the
evolution seen in this plot may also explain why the slope of the $L_V
- \sigma_0$ relation for the Local Group GCs is steeper than that
observed for the youngest YMCs.

In summary, the results of $N$-body simulations show that -- for
clusters in relatively quiescent environments -- the tracks followed
in the $L_V - \sigma_0$ plane are broadly similar to those shown in
Fig.~\ref{diagnostic1.fig}. In fact, the evolution toward smaller
$\sigma_0$ seen in Fig.~\ref{fig:nbody_evol} suggests, that at late
times the surviving YMCs will tend to lie closer to the Local Group GC
relation than the YMCs aged to a fiducial age of 12 Gyr do in
Fig.~\ref{diagnostic1.fig}.

\section{Discussion}
\label{discussion.sec}

\subsection{Photometric versus dynamical mass estimates}

Thus far, we have considered the location of our sample of
extragalactic YMCs in the two-dimensional $L_V-\sigma_0$
projection. However, for any virialised system, we can look for a
fundamental plane akin to that of elliptical galaxies and spiral
bulges, using size as a third parameter. If a cluster's M/L ratio is
constant across its volume, and the projected half-light radius
satisfies $R_{hp} = \frac{3}{4} R_h$ (applicable for most realistic
cluster profiles; Spitzer 1987) and therefore represents the half-mass
radius, we can relate the cluster's mass to its velocity dispersion
via the virial theorem (Spitzer 1987):
\begin{equation}
M_{\rm dyn} \approx 10 \frac{\sigma_{\rm obs}^2 R_{hp}}{G}.
\label{virial.eq}
\end{equation}
Here, $\sigma_{\rm obs}$ is the observed total velocity dispersion of
the cluster.

In view of the uncertainties in the IMF discussed in the previous
section, we have calculated the photometric masses of all of our
sample clusters using four different IMF descriptions and SSP models
computed for the relevant observational bandpasses. The results are
presented in Table \ref{masses.tab}. Where available, uncertainties
are based on the maximum uncertainties in the fundamental parameters
determining the exact conversion from luminosities to masses (such as
uncertainties in the YMC ages, photometry, or extinction values). We
have also included in Table \ref{masses.tab} previously published
photometric mass estimates, as well as estimates of the clusters'
dynamical masses based on the observed velocity dispersions and
half-light radii.

\begin{table*}
\caption[ ]{\label{masses.tab}Cluster mass estimates, using a variety of estimators.\\
The nomenclature used for the mass estimators is of the form ``SSP
models''--``IMF prescription'', as explained in the text.}
{\scriptsize
\begin{center}
\hspace*{-1.3cm}
\begin{tabular}{lcrcccccr}
\hline
\hline
\multicolumn{1}{c}{Cluster} & \multicolumn{1}{c}{$M_{\rm phot,lit.}$}
& \multicolumn{1}{c}{Ref.}  & \multicolumn{1}{c}{GALEV--Salpeter} &
\multicolumn{1}{c}{SB99--Salpeter} & \multicolumn{1}{c}{GALEV--KTG93}
& \multicolumn{1}{c}{GALEV--Kroupa01} & \multicolumn{1}{c}{$M_{\rm
dyn}$} & \multicolumn{1}{c}{Ref.} \\
\cline{4-8}
& \multicolumn{1}{c}{(M$_\odot$)} & & \multicolumn{6}{c}{(M$_\odot$)} \\
\hline
Antennae-$\mbox{[WS95]331}$ &                & & $(3.8\pm0.6)\hfill\times 10^4$ & $(4.3\pm1.3)\hfill\times 10^4$ & 
                $(2.6\pm0.2)\hfill\times 10^5$ & $(2.1\pm0.1)\hfill\times 10^5$ & $(0.52\pm0.2)\hfill\times 10^6$& 16 \\
Antennae-$\mbox{[WS95]355}$ &                & & $(3.8\pm0.3)\hfill\times 10^4$ & $(2.4\pm0.2)\hfill\times 10^4$ & 
                $(2.1\pm0.3)\hfill\times 10^5$ & $(1.7\pm0.3)\hfill\times 10^5$ & $(4.7\pm0.6)\hfill\times 10^6$ & 15 \\
Antennae-$\mbox{[W99]1}$    &                & & $5.8        \hfill\times 10^5$ & $3.8        \hfill\times 10^5$ &
                $4.0        \hfill\times 10^6$ & $3.2        \hfill\times 10^6$ & $(6.5\pm1.2)\hfill\times 10^5$ & 15 \\
Antennae-$\mbox{[W99]2}$    &                & & $6.6        \hfill\times 10^5$ & $2.1        \hfill\times 10^5$ &
                $3.9        \hfill\times 10^6$ & $3.2        \hfill\times 10^6$ & $(2.0\pm0.2)\hfill\times 10^6$ & 15 \\
Antennae-$\mbox{[W99]15}$   &                & & $1.4        \hfill\times 10^5$ & $1.2        \hfill\times 10^5$ &
                $7.9        \hfill\times 10^5$ & $7.1        \hfill\times 10^5$ & $(3.3\pm0.5)\hfill\times 10^6$ & 15 \\
Antennae-$\mbox{[W99]16}$   &                & & $1.7        \hfill\times 10^5$ & $1.8        \hfill\times 10^5$ &
                $9.5        \hfill\times 10^5$ & $7.3        \hfill\times 10^5$ & $(3.2\pm0.5)\hfill\times 10^6$ & 15 \\
Antennae-$\mbox{[M03]}$     &                & & $4.3        \hfill\times 10^5$ & $2.6        \hfill\times 10^5$ &
                $3.0        \hfill\times 10^6$ & $2.4        \hfill\times 10^6$ & $(0.85\pm0.2)\hfill\times 10^6$& 16 \\
IC  342-NC    &                              & & $(2.5-9.7)  \hfill\times 10^5$ & $(0.6-5.0)  \hfill\times 10^5$ & 
                $(0.8-1.2)  \hfill\times 10^6$ & $(0.7-1.5)  \hfill\times 10^6$ & $(6.0\pm2.4)\hfill\times 10^6$ &  3 \\
M82-F         &                              & & $(7.5\pm1.7)\hfill\times 10^6$ & $(3.9\pm0.9)\hfill\times 10^6$ & 
                $(1.1\pm0.3)\hfill\times 10^7$ & $(1.2\pm0.4)\hfill\times 10^7$ & $(1.2\pm0.1)\hfill\times 10^6$ & 20 \\
& & & & & & & $(6.6 \pm 0.9)\hfill\times 10^5$ & 13$^a$\\
& & & & & & & $(7.0 \pm 1.2)\hfill\times 10^5$ & 13$^a$\\
M82 MGG-9     &                              & & $(2.7\pm1.0)\hfill\times 10^6$ & $(2.0\pm1.3)\hfill\times 10^6$ &
                $(1.2-1.4)  \hfill\times 10^7$ & $(0.9-1.3)  \hfill\times 10^7$ & $(1.5\pm0.3)\hfill\times 10^6$ & 12 \\
M82 MGG-11    &                              & & $(1.6\pm0.6)\hfill\times 10^6$ & $(1.2\pm0.8)\hfill\times 10^6$ &
                $(6.8-8.3)  \hfill\times 10^6$ & $(5.4-7.6)  \hfill\times 10^6$ & $(3.5\pm0.7)\hfill\times 10^5$ & 12 \\
NGC 1042-NC   &                              & & $6.7        \hfill\times 10^6$ & $3.8        \hfill\times 10^6$ &  
                $3.2        \hfill\times 10^6$ & $4.2        \hfill\times 10^6$ & $3.0        \hfill\times 10^6$ &  1 \\
NGC 1487-1    &                              & & $2.7        \hfill\times 10^5$ & $1.5        \hfill\times 10^5$ &
                $1.9        \hfill\times 10^6$ & $1.6        \hfill\times 10^6$ & $(1.5\pm0.2)\hfill\times 10^6$ & 16 \\
NGC 1487-2    &                              & & $2.3        \hfill\times 10^5$ & $1.3        \hfill\times 10^5$ &
                $1.6        \hfill\times 10^6$ & $1.3        \hfill\times 10^6$ & $(1.0\pm0.2)\hfill\times 10^6$ & 16 \\
NGC 1487-3    &                              & & $1.2        \hfill\times 10^5$ & $5.7        \hfill\times 10^4$ &
                $8.5        \hfill\times 10^5$ & $6.8        \hfill\times 10^5$ & $(2.3\pm0.2)\hfill\times 10^6$ & 16 \\
NGC 1569-A1   & $(1.1-2.1)\hfill\times 10^6$ &  2 & $(1.2-14.6) \hfill\times 10^5$ & $(2.1-7.6)  \hfill\times 10^5$ &
                $(2.0-4.0)  \hfill\times 10^6$ & $(1.3-4.0)  \hfill\times 10^6$ & $(3.3\pm0.5)\hfill\times 10^5$ &  6 \\
              & & & & & & & $2.8\hfill\times 10^5$ &  4 \\
              & & & & & & & $8.3\hfill\times 10^5$ &  5 \\
NGC 1614-NC1  &                              & & $6.8        \hfill\times 10^8$ & $(1.7-3.6)  \hfill\times 10^8$ &
                $2.1        \hfill\times 10^9$ & $2.2        \hfill\times 10^9$ & $1.6        \hfill\times 10^9$ & $19^b$ \\
NGC 1614-NC2  &                              & & $7.4        \hfill\times 10^8$ & $(1.6-3.9)  \hfill\times 10^8$ &
                $2.3        \hfill\times 10^9$ & $2.4        \hfill\times 10^9$ & $1.6        \hfill\times 10^9$ & $19^b$ \\
NGC 1705-I$^c$& $7\hfill\times 10^6$ & 14 & $(2.6\pm1.0)\hfill\times 10^6$ & $(1.5\pm0.2)\hfill\times 10^6$ &
                $(1.0\pm0.1)\hfill\times 10^7$ & $(9.7\pm0.9)\hfill\times 10^6$ & $(8.2\pm2.1)\hfill\times 10^4$ & 7 \\
              & $1.5\hfill\times 10^6$ & 17 \\
              & $2.5\hfill\times 10^5$ & 18 \\
NGC 4214-10   &                              & & $2.9^{+0.3}_{-0.6}\hfill\times 10^5$ & $(1.6\pm0.3)\hfill\times 10^5$ & 
                $2.4^{+0.1}_{-0.2}\hfill\times 10^5$ & $3.1^{+0.2}_{-0.3}\hfill\times 10^5$ & $(2.6\pm1.0)\hfill\times 10^5$ & 9 \\
NGC 4214-13   &                              & & $1.1^{+0.1}_{-0.2}\hfill\times 10^6$ & $(6.1\pm1.0)\hfill\times 10^5$ & 
                $9.2^{+0.2}_{-0.8}\hfill\times 10^5$ & $(1.2\pm0.1)\hfill\times 10^6$ & $(1.48\pm0.24)\hfill\times 10^6$ & 9 \\
NGC 4449-27   &                              & & $4.0^{+1.7}_{-1.3}\hfill\times 10^5$ & $(2.3\pm0.8)\hfill\times 10^5$ & 
                $2.0^{+0.3}_{-0.4}\hfill\times 10^5$ & $3.0^{+0.7}_{-0.6}\hfill\times 10^5$ & $(2.1\pm0.9)\hfill\times 10^5$ & 9 \\
NGC 4449-47   &                              & & $5.5^{+0.9}_{-0.6}\hfill\times 10^5$ & $(3.2\pm0.5)\hfill\times 10^5$ & 
                $4.0^{+0.3}_{-0.1}\hfill\times 10^5$ & $5.6^{+0.5}_{-0.4}\hfill\times 10^5$ & $(4.6\pm1.6)\hfill\times 10^5$ & 9 \\
NGC 5236-502  & $(4.49\pm0.86)\hfill\times 10^5$ & $10^d$ & $7.0^{+1.2}_{-0.8}\hfill\times 10^5$ & $(3.7\pm0.6)\hfill\times 10^5$ &
                $7.9^{+0.4}_{-0.3}\hfill\times 10^5$ & $9.7^{+0.8}_{-0.5}\hfill\times 10^5$ & $(5.15\pm0.83)\hfill\times 10^5$ & 10 \\
              & $(6.56\pm1.26)\hfill\times 10^5$ & $10^d$ \\
NGC 5236-805  & $(1.93\pm1.42)\hfill\times 10^5$ & $10^d$ & $2.5^{+2.2}_{-1.3}\hfill\times 10^5$ & $1.9^{+0.1}_{-1.3}\hfill\times 10^5$ &
                $1.0^{+0.1}_{-0.2}\hfill\times 10^6$ & $8.9^{+1.8}_{-2.6}\hfill\times 10^5$ & $(4.16\pm0.67)\hfill\times 10^5$ & 10 \\
              & $(2.84\pm2.06)\hfill\times 10^5$ & $10^d$ \\
NGC 6946-1447 & $(5.5-8.2)\hfill\times 10^5$ & 8 & $1.6^{+0.3}_{-0.7}\hfill\times 10^6$ & $1.3^{+0.3}_{-0.4}\hfill\times 10^6$ & 
                $5.9^{+0.5}_{-0.6}\hfill\times 10^6$ & $(5.2\pm0.9)\hfill\times 10^6$ & $(1.8\pm0.5)\hfill\times 10^6$ &  8,9 \\
NGC 7252-W3   & $(4.0-7.2)\hfill\times 10^7$ & 11 & $8.8        \hfill\times 10^7$ & $5.1        \hfill\times 10^7$ &
                $6.3        \hfill\times 10^7$ & $8.7        \hfill\times 10^7$ & $(8\pm2)    \hfill\times 10^7$ & 11 \\
\hline
\end{tabular}
\end{center}
{\sc Notes:} $^a$ based on $H$ and $I$-band spectroscopy (first and
second line, respectively); $^b$ based on barycentric motions; $M_{\rm
dyn} = 2 \times 10^8 M_\odot$ if virialised; $^c$ The differences
among the existing photometric mass estimates are mostly caused by
varying distance estimates to the galaxy (Ho \& Filippenko 1996b);
$^d$ The photometric mass estimates are for a Kroupa01 and a Salpeter
IMF, covering masses down to 0.1 M$_\odot$ (first and second line,
respectively). {\bf References:} 1, this work, based on data from
B\"oker et al. (2004, 2005); 2, Anders et al. (2004); 3, B\"oker et
al. (1999); 4, de Marchi et al. (1997); 5, Gilbert \& Graham (2001);
6, Ho \& Filippenko (1996a); 7, Ho \& Filippenko (1996b); 8, Larsen et
al. (2001); 9, Larsen et al. (2004); 10, Larsen \& Richtler (2004);
11, Maraston et al. (2004); 12, McCrady et al. (2003); 13, McCrady et
al. (2005); 14, Melnick et al. (1985); 15, Mengel et al. (2002); 16,
Mengel (2003); 17, Meurer et al. (1992); 18, Meurer et al. (1995); 19,
Puxley \& Brand (1999); 20, Smith \& Gallagher (2001).}
\end{table*}

\begin{figure}
\psfig{figure=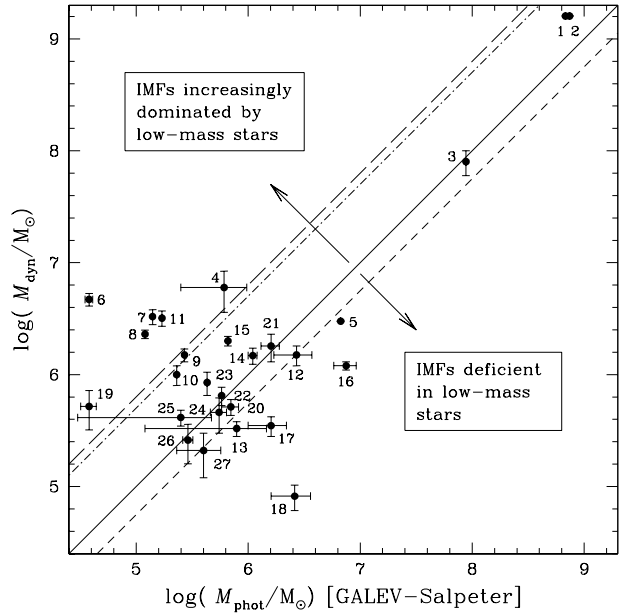,width=8.7cm}
\caption[]{Comparison of photometric with dynamical mass estimates for
the young massive star clusters analysed in this paper. The solid line
represents the loci of clusters of which the dynamical mass is exactly
reproduced by a Salpeter IMF covering masses from 0.1 to 100
M$_\odot$, using the GALEV SSP models. The short-dashed line
represents photometric masses for a Salpeter IMF truncated below 1
M$_\odot$, using the Starburst99 SSP models, while the long-dashed and
dot-dashed lines correspond to the photometric mass estimates obtained
using the KTG93 and Kroupa01 IMFs, respectively, again using the GALEV
SSP models.}
\label{masscf.fig}
\end{figure}

Figure \ref{masscf.fig} provides a projection of the ``YMC fundamental
plane'' defined in the space of the YMCs' luminosities, velocity
dispersions and sizes. We show the distribution of our sample YMCs in
the plane defined by the photometric vs. the dynamical mass estimates;
the photometric mass estimates are based on converting the cluster
luminosities to masses using the GALEV SSPs under the assumption of a
Salpeter IMF from 0.1 to 100 M$_\odot$. The solid line of equality
represents the loci where our sample clusters would be found if they
were characterised by this Salpeter IMF, and a constant M/L ratio
throughout. The other lines, offset from the solid line, are
calculated for the alternative IMFs considered for the photometric
mass estimates listed in Table \ref{masses.tab}. We can conclude that
most of our sample YMCs are scattered closely around the line of
equality, which provides additional evidence that they are
characterised by IMFs (or present-day MFs) similar to the standard
Salpeter IMF. Only few objects, including the M82 clusters F and
MGG-11, and NGC 1705-I, are found in the region where we expect to see
the effects of either a low-mass cut-off or significant mass
segregation. This lends support to McCrady et al.'s (2003, 2005)
suggestion that these M82 clusters are affected by significant
primordial mass segregation, and suggests a similar effect for NGC
1705-I. In this context, we note that the straightforward application
of the virial theorem, Eq. (\ref{virial.eq}), which is based on a
single-mass model for all stars contained in the system, tends to
underestimate a system's dynamical mass by a factor of $\sim 2$
compared to more realistic multi-mass models (e.g., Mandushev et
al. [1991], based on an analysis of the observational
uncertainties). This effect potentially reduces the number of clusters
in Fig. \ref{masscf.fig} scattered toward MFs defined by low-mass
cut-offs or YMCs dominated by significant mass segregation even
further.

\subsection{Implications}
\label{implications.sec}

The origin of the tight relationship between the absolute magnitude
and central velocity dispersion for all Local Group GCs remains an
unsolved puzzle. Djorgovski (1991, 1993; see also Djorgovski \& Meylan
1994) suggested that the relation evolved from a primordial scaling
relation, $m/M_\odot \propto \sigma$ (assuming a constant M/L ratio
among GCs), which would be subsequently altered by tidal shocks,
leading to mass (and therefore luminosity) losses. This would be more
efficient for the less massive clusters, thus resulting in a
steepening of the relationship to its currently observed
form. McLaughlin (2003) suggests that the relation is linked to the
mass-dependent star-formation efficiencies in giant molecular clouds,
the progenitors of star clusters. We note that the fact that {\it all}
Local Group GCs are found scattering closely around the relationship
implies that its origin must be related to GC-internal processes. The
tightness of the relationship rules out significant environmental
effects as principal cause for its origin. This is simply because the
Local Group GCs are found in a wide variety of environments, ranging
from the high-density environments in the Galaxy and M31, via the
intermediate density operating in M33, to the (very) low density
environments in the dwarf satellite galaxies (LMC, SMC, Fornax
dSph). A similar conclusion was reached by McLaughlin (2000a) when he
noted that Galactic GCs at larger Galactocentric distances exhibit a
smaller scatter about the relationship than those closer to the Milky
Way.

The fact that we find that our sample YMCs, when evolved to a common
age of 12 Gyr using the Salpeter IMF, may also evolve to loci close to
the best-fitting GC relationship implies that the initial conditions
governing these YMC must have been very similar to those responsible
for the formation of the old Local Group GCs. This, therefore,
provides an argument in favour of the suggestion that most of these
YMCs may in fact be proto-GCs. It also suggests that a large number of
the present-day young compact LMC (and SMC) clusters, as well as the
large majority of the Galactic open clusters, all of which are
currently found to occupy regions close to the old GC relationship (in
some cases further toward fainter magnitudes than any of the known
GCs, for a given central velocity dispersion), are unlikely to survive
until they reach GC-type ages of $\gtrsim 10$ Gyr.

Thus far, we have been dealing predominantly with internal cluster
processes that might prevent (a number of) the YMCs from surviving for
a Hubble time. The most likely {\it internal} processes leading to
cluster disruption were found to be related to variations in the
IMF. However, we note that our predictions for the future fate of our
sample clusters should only be adopted as first-order
approximations. Until now, we have only mentioned external disruptive
effects in passing, and have assumed our clusters to reside in
quiescent galactic disc environments. This assumption is clearly not
justified in a number of cases considered in this paper.

One should realise that star cluster survivability also -- and
crucially so -- depends on external factors affecting its stellar
content, such as tidal shocking by galactic discs, bulges, spiral arms
and giant molecular clouds (GMCs), and the associated ram-pressure
stripping. These external effects will accelerate the cluster
disruption time-scale relative to that caused by cluster-internal
effects.

In a recent study, Boutloukos \& Lamers (2003) derived an empirical
expression for the ``characteristic'' cluster disruption time-scale
(i.e., the time-scale on which a $10^4$ M$_\odot$ cluster will
dissolve, assuming instantaneous disruption), and found that -- for a
given cluster system and environment -- this time-scale is entirely
dependent on the initial mass of the cluster, as $t_{\rm dis} \propto
(M_{\rm cl}/10^4 {\rm M}_\odot)^{0.60\pm0.02}$ (see also Lamers,
Gieles \& Portegies Zwart 2005, who confirmed this prediction using
{\it N}-body simulations). Boutloukos \& Lamers (2003) derived
characteristic cluster disruption time-scales for the cluster systems
in the solar neighbourhood, the SMC, and in selected regions of M33
and the interacting galaxy M51. In de Grijs et al. (2003a,c), we
extended this sample to include the fossil starburst region M82 B, and
the interacting systems NGC 3310 and NGC 6745.

In de Grijs et al. (2003c), we concluded that the very short
characteristic cluster disruption time-scale for the clusters in M82 B
is most likely caused by the very high ambient density of its
interstellar medium (ISM), leading to cluster disruption on similarly
short time-scales as in the high-density centre of M51.\footnote{For
counterarguments see Mengel et al. (2002), who explained the unusual
M/L ratios found for the YMCs in the overlap region between the
merging galaxies in the Antennae system by suggesting that higher
ambient pressures might be conducive to the formation of more low-mass
stars, leading to more stable clusters.}

If we place our own results in this context, we see that four of the
six clusters that are expected to evolve to beyond the $3\sigma_{\rm
scatter}$ boundary by an age of 12 Gyr are in fact located in the
high-density overlap region in the Antennae galaxies. We would expect
these objects to dissolve on shorter-than-average time-scales, simply
because of the higher density ISM in which they are embedded, and
because of the high pressure and tidal shocks expected in the ongoing
merger. Similarly, the remaining two objects (NGC 1487-3 and IC
342-NC) are located in high-density galactic centre environments. By
the same token, NGC 1487-1 and 2, and NGC 5236-805 are located in
similarly high-density environments; their luminosity evolution arrows
do, in fact, overshoot the $2\sigma_{\rm scatter}$ envelope. This is
supported by a recent study by Lamers et al. (2005), based on
numerical simulations. We caution that the results for the NGC 1487
clusters should be treated with caution in view of the large
photometric uncertainties caused by the passband conversion
applied. However, if we take the evolution of the central velocity
dispersion into account, all of these objects may well evolve to loci
within the $2 \sigma_{\rm scatter}$ boundary by the time they age to
12 Gyr.

If we assume that the {\it initial} MF of all of these objects was
roughly constant for the entire YMC sample, this implies that tidal
effects and their location in regions of higher-than-average density
must have affected the stellar content of these clusters already on
time-scales as short of $\sim 10^7-10^8$ yr, i.e., a significant
fraction of the low-mass stars in these objects has likely been
tidally stripped already during their very short lifetimes. Ongoing
tidal effects would lead to luminosity evolution to still fainter
magnitudes than implied by assuming a Salpeter-type IMF.

Now that we have established that a number of our sample clusters are
already likely to have been affected significantly by tidal effects
and externally induced disruption, despite their young ages, we return
to the origin of the tight GC relation. With the remainder of our
sample YMCs, except the most massive objects that may be governed by
the FJ relation rather than the old GC correlation, we can now test
the suggestion by Djorgovski (1991, 1993) and Djorgovski \& Meylan
(1994) that at the time of {\it proto-globular} cluster formation the
clusters' (central) velocity dispersion correlated linearly with their
luminosity. For the following arguments, one needs to keep in mind
that we have shown (i) that the remainder of our YMC sample shows
behaviour consistent with their stellar content being described by a
Salpeter-type present-day (and presumably initial) MF (assuming that
they are to obey the $L_V-\sigma_0$ relationship at old age), (ii)
that all of these clusters are likely governed by a very similar IMF,
and (iii) that they are possible GC progenitors, in the absence of
significant external disruptive processes.

\begin{figure}
\psfig{figure=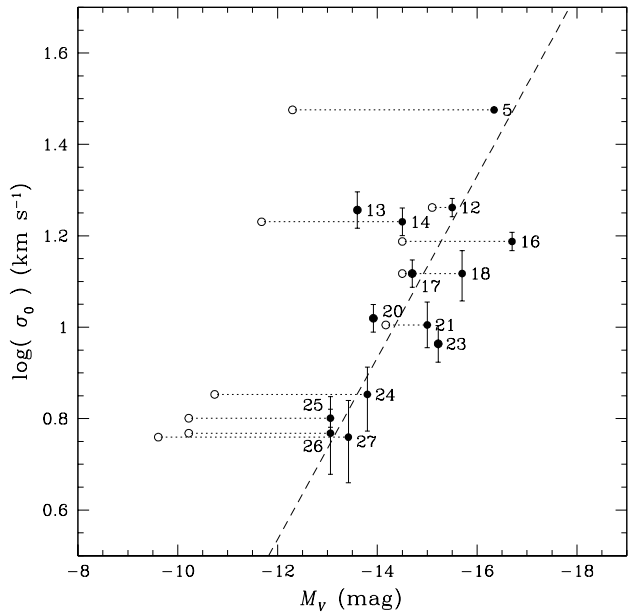,width=8.7cm}
\caption[]{Diagnostic $L_V-\sigma_0$ diagram for our sample YMCs that
have likely not (yet) been significantly affected by external tidal
forces, evolved back in time to a common age of 8 Myr (filled
circles). We show the evolutionary correction from their present-day
loci (open circles) by means of the dashed lines. The clusters are
numbered following Table \ref{clusids.tab}. The dashed line is the
best-fitting relationship to the 8 Myr-old YMC sample. Error bars have
been included where available}
\label{initconds.fig}
\end{figure}

With this picture in mind, we can now evolve the present-day
luminosities of these YMCs back to a common age corresponding to the
youngest age found in this cluster sample, i.e., 8 Myr, again using
the GALEV SSPs with a standard Salpeter IMF. We show the results of
this exercise in Fig. \ref{initconds.fig}.

For the first time, we can now assess the almost-initial conditions of
proto-GCs in our diagnostic $L_V-\sigma_0$ diagram.  The best-fitting
(dashed) relationship corresponds to
\begin{equation}
\sigma_0 (\mbox{km s}^{-1}) \propto \Bigl( \frac{L_V}{L_\odot}
\Bigr)^{0.48 \pm 0.10}
\end{equation}
or
\begin{equation}
\frac{L_V}{L_\odot} \propto \sigma_0^{2.1^{+0.5}_{-0.4}} (\mbox{km
s}^{-1}),
\end{equation}
with correlation coefficient $\Re = -0.71$, when expressed in
logarithmic units. This result excludes a linear $L_V \propto \sigma$
relation at the $\gtrsim 2.5 \sigma$ level. The exact relationship is
somewhat dependent on the exact functional form of the IMF
adopted. For instance, if we had adopted a Kroupa01 IMF, the exponents
in Eq. (4) and (5) would have been $0.34 \pm 0.08$ and
$2.9^{+0.0}_{-0.5}$, respectively.

This relation can be understood in terms of the state of equilibrium
of the observed clusters. For a cluster in virial equilibrium we have
(see, e.g., Binney \& Tremaine 1987)
\begin{equation}
\label{sigma.eq}
\sigma^2 \approx 0.4\frac{{\rm G}\Upsilon L}{r_{\rm h}} \, ,
\end{equation}
assuming that the cluster has a constant M/L ratio, $\Upsilon$. Thus,
the observed relation for the youngest YMCs has exactly the form
expected for clusters in virial equilibrium, provided that (i) the
cluster radii are independent of their luminosities, (ii) the cluster
radii have not changed significantly since the clusters were $8$ Myr
old, and (iii) the ratio of central velocity dispersion $\sigma_0$ to
the total cluster dispersion is independent of luminosity. With regard
to the first point, McLaughlin (2000a) found that the half-light radii
of the Milky Way clusters are indeed independent of their total
masses. Similarly, Harris et al. (2002) found no significant
correlation between cluster sizes and their absolute magnitudes in a
sample of clusters surrounding the giant elliptical galaxy NGC 5128,
and neither did we find any such correlation between the half-light
radii and absolute magnitudes of our Local Group GC sample. Note that
in the $N$-body simulations presented in Section \ref{sigma.sec} the
low-mass clusters were systematically smaller in radius than the more
massive clusters, which is why the simulated low-mass clusters do not
lie on the young YMC relation (see Fig. \ref{fig:nbody_evol}). The
half-light radius of a cluster is most significantly affected by the
expulsion of gas immediately following the end of star formation,
which results in the expansion of the cluster by up to a factor of
$4-5$ (Boily \& Kroupa 2003, Goodwin 1997b). Bound clusters rapidly
re-establish equilibrium. It is therefore reasonable to expect that
the half-light radii have not evolved significantly since an age of 8
Myr -- even if some of the clusters have expanded since that time, Eq.
(\ref{sigma.eq}) shows that the magnitude of this effect will be less
than 0.35 dex in $\log\sigma_0$. Finally, the absence of significant
luminosity dependence of the ratio of central to total cluster
velocity dispersions is expected for clusters in
equilibrium. Fig. \ref{initconds.fig} is thus consistent with the
youngest YMCs having rapidly achieved virial equilibrium. In order to
strengthen this result, it would be interesting to use accurate
determinations of cluster radii to confirm the independence of the
cluster sizes and luminosities in the extragalactic YMC sample.

The simple, virial $L_V-\sigma_0$ relation for the youngest clusters
in our YMC sample may be the pre-cursor for the fundamental plane of
globular clusters. Clearly, quiescent evolution would be expected to
transform a primordial linear relation into another linear relation
since two clusters which are initially close together in the $L_V -
\sigma_0$ plane will evolve similarly provided their external
environments do not differ too greatly. The change in the slope of the
relation is then probably due to the dependence of the $\sigma_0$
evolution on the mass of the cluster. The increased relaxation time of
more massive clusters would be expected to lead to less evolution in
these clusters than is seen in the lower mass clusters; we also note
that the amount of luminosity evolution is driven by relative age
differences and, to first order, independent of a cluster's initial
luminosity. This would naturally account for the steeper slope of the
late-time relation seen for the Local Group GCs. Further numerical
simulations are required to confirm that this picture is consistent in
all respects with the observations.

\section{Summary and conclusions}
\label{summary.sec}

In this paper, we have presented a new analysis of the properties and
possible evolutionary paths of the YMCs forming profusely in intense
starburst environments, such as those associated with galaxy
interactions and mergers. The method hinges on the empirical
relationship for old Galactic and M31 GCs, which occupy a tightly
constrained locus in the plane defined by their $V$-band luminosities,
$L_V$ (or, equivalently, absolute magnitudes, $M_V$) and central
velocity dispersions, $\sigma_0$ (Djorgovski et al. 1997, McLaughlin
2000a, and references therein).

We added to the Galactic and M31 GC sample the old compact Magellanic
Cloud clusters, and the M33 and Fornax dSph GCs for which the relevant
observational parameters were available in the literature. The
relationship between $L_V$ and $\sigma_0$ for this increased GC
sample, $L_V/L_\odot \propto \sigma_0^{1.57 \pm 0.10}$ (km s$^{-1}$),
is within the uncertainties consistent with Djorgovski et al.'s (1997)
determination for the smaller Galactic and M31 GC sample. The
tightness of the relationship for a sample drawn from environments as
diverse as those found in the Local Group, ranging from high to very
low ambient densities, implies that its origin must be sought in
intrinsic properties of the GC formation process itself, rather than
in external factors. This is further supported by McLaughlin's (2000a)
result that GCs at greater Galactocentric distances exhibit a smaller
scatter about the relation than closer objects.

Encouraged by the tightness of the GC relationship, we also added the
available data points for the YMCs in the local Universe, including
nuclear star clusters, for which velocity dispersion information was
readily available. In order to be able to compare them to the
ubiquitous old Local Group GCs, we evolved their luminosities to a
common age of 12 Gyr, adopting the ``standard'' (solar neighbourhood)
Salpeter IMF covering masses from 0.1 to 100 M$_\odot$, and assuming
stellar evolution as described by the GALEV SSPs. Based on a careful
assessment of the uncertainties associated with this luminosity
evolution, we concluded that the most important factor affecting the
robustness of our conclusions is the adopted form of the stellar IMF.

We found that if we adopt the Salpeter IMF as the basis for the YMCs'
luminosity evolution, the large majority will evolve to loci within
twice the observational scatter around the best-fitting GC
relationship (although systematically to somewhat fainter
luminosities). Using more realistic IMF descriptions, our YMC sample
do, in fact, end up scattering more closely about the improved Local
Group GC relationship. In the absence of significant external
disturbances, this implies that these objects may potentially survive
to become old GC-type objects by the time they reach a similar
age. Thus, these results provide additional support to the suggestion
that the formation of proto-GCs appears to be continuing until the
present, a conclusion we reached independently based on the
statistical treatment of the $\sim 1$ Gyr-old intermediate-age star
cluster system in M82's fossil starburst region B (de Grijs et
al. 2003b). Detailed case by case comparisons between our results
based on this new method with those obtained previously and
independently based on dynamical mass estimates and M/L ratio
considerations lend significant support to the feasibility and
robustness of our new method, and provide a key insight into the
inherent uncertainties associated with any of the methods used in this
field. The key characteristic and main advantage of this method
compared to the more complex analysis involved in using dynamical mass
estimates for this purpose is its simplicity and empirical
basis. Where dynamical mass estimates require one to obtain accurate
size estimates and to make assumptions regarding a system's virialised
state and M/L ratio, these complications can now be avoided by using
the empirically determined GC relationship as reference. The only
observables required are the system's (central or line-of-sight)
velocity dispersion and photometric properties. McLaughlin (2000a) has
shown that this is, in fact, a physically relevant correlation, since
(i) the $E_{\rm b}, L$ diagram (where $E_{\rm b}$ is the cluster
binding energy) is composed of physically meaningful quantities, and
(ii) the scatter about the correlation is of the same order as the
observational uncertainties.

Careful analysis of those YMCs that would overshoot the GC
relationship significantly if they were to survive for a Hubble time
(and are characterised by a Salpeter-type initial or present-day MF)
showed that their unusually high ambient density has probably already
had a significant effect on their stellar content, despite their young
ages, thus altering their present-day MF in a such a way that they
have become unable to survive for any significant length of
time. This is, again, supported by independent analyses, thus further
strengthening the robustness of our new approach. The expected loci in
the $L_V-\sigma_0$ plane that these objects would evolve to over a
Hubble time are well beyond any GC luminosities for a given velocity
dispersion, leading us to conclude that they will either dissolve long
before reaching GC-type ages, or that they must be characterised by a
present-day MF that is significantly depleted in low-mass stars (or
highly mass segregated), thus also resulting in fast dispersion. This,
therefore, allows us to place moderate limits on the functionality of
their present-day MFs.

In order to investigate whether dynamical evolution would have a
dramatic impact on the evolution of clusters in the $L_V - \sigma_0$
plane, we analysed the results of a number of $N$-body simulations.
The velocity dispersions of the model clusters were calculated in a
manner analogous to that used for the observed clusters. We concluded
that the evolution of the observed $\sigma_0$ is relatively small for
clusters that survive to old age, and thus our conclusions remain
unchanged.

Based on our analysis of the objects with the largest velocity
dispersions, including the nuclear star clusters, we conclude that the
recently discovered UCDs in the Fornax cluster may be most closely
related to stripped dSph or dE nuclei. We also show that the unusual
Galactic GC NGC 2419 is unlikely to be a similar type of object,
despite recent suggestions to the contrary.

Finally, we evolved those YMCs that appear to be least affected by
external disruptive effects and are likely to be well-represented by
Salpeter-type IMFs back to a common young age of 8 Myr, in order to
assess the $L_V-\sigma_0$ relationship in almost-initial
conditions. The resulting best-fitting relationship, $L_V/L_\odot
\propto \sigma_0^{2.0^{+0.5}_{-0.4}}$ (km s$^{-1}$), implies that
these clusters follow a simple virial relation. The evolution of
relatively undisturbed star clusters in the $L_V - \sigma_0$ plane, as
seen in our $N$-body simulations, will subsequently transform this
relation into the steeper relation displayed by the Local Group
GCs. The existence of a simple, virial $L_V-\sigma_0$ relationship for
the youngest YMCs may therefore constitute the origin of the GC
fundamental plane.

\section*{Acknowledgments} RdG acknowledges the stimulating atmosphere 
at the 2004 Guillermo Haro workshop at INAOE, Tonantzintla, Mexico,
during which much of this work was done; he also thanks the Royal
Society for providing travel funding to attend this workshop. We are
very grateful to Jarrod Hurley for providing us with data from his
$N$-body simulations in advance of publication. We thank Peter Anders
for re-calculating his GALEV SSP models for the range of IMF
representations discussed in this paper, and acknowledge stimulating
discussions with Roberto Terlevich and Jay Gallagher, and a number of
insightful comments by the anonymous referee. MIW thanks Gijs Nelemans
for valuable discussions and PPARC for financial support. This
research has made use of the SIMBAD database, operated at CDS,
Strasbourg, France, and of the WEBDA database maintained by
Jean-Claude Mermilliod at http://obswww.unige.ch/webda/.

\end{document}